\documentclass{aa}  

\usepackage{graphicx}
\usepackage{xcolor}
\usepackage{txfonts}
\usepackage[colorlinks=true]{hyperref}
\hypersetup{linkcolor=blue,citecolor=blue,filecolor=blue,urlcolor=blue}
\begin{document} 

   \title{Dark and bright sides of the Broad Line Region clouds as seen in the FeII emission of SDSS RM 102}


  \author{Alberto Floris \inst{1, 8}
  \and
  Ashwani Pandey \inst{1,2}\thanks{PIFI visiting scientist}
  \and
  Bozena Czerny \inst{1}
     \and
  Mary Loli Martinez Aldama \inst{3,4,5}
      \and
  Swayamtrupta Panda \inst{6, 7, \thanks{Gemini Science Fellow}}
     \and
  Paola Marziani \inst{8} 
    \and
    Raj Prince \inst{1,9}}

        \institute{Center for Theoretical Physics, Polish Academy of Sciences, Al. Lotnik\'ow 32/46, 02-668 Warsaw, Poland\\
        \and
        Key Laboratory for Particle Astrophysics, Institute of High Energy Physics, Chinese Academy of Sciences, 19B Yuquan Road, Beijing 100049, P. R. China\\
        \and
        Astronomy Department, Universidad de Concepción, Barrio Universitario S/N, Concepción 4030000, Chile\\
        \and
        Millennium Institute of Astrophysics MAS, Nuncio Monse\~nor Sotero Sanz 100, Of. 104, Providencia, Santiago, Chile\\
        \and
        Millennium Nucleus on Transversal Research and Technology to Explore Supermassive Black Holes (TITANS), Chile\\
        \and
        International Gemini Observatory/NSF NOIRLab, Casilla 603, La Serena, Chile\\
        \and
        Laborat\'{o}rio Nacional de Astrofísica, MCTI, R. dos Estados Unidos 154, Na\c{c}\~{o}es, CEP 37504-364, Itajub\'a, Brazil\\
        \and 
        National Institute for Astrophysics (INAF), Astronomical Observatory of Padova, IT-35122 Padova, Italy \\
        \email{alberto.floris@inaf.it}
        \and
        Department of Physics, Institute of Science, Banaras Hindu University, Varanasi-221005, India
        }

   \date{}

 
  \abstract
   {Contamination from singly ionized iron emission is one of the greatest obstacles to determining the intensity of emission lines in the UV and optical wavelength ranges.} 
   {This study presents a comprehensive analysis of the {\rm Fe{\sc ii}} emission in the bright quasar RM 102, based on the most recent version of the {\tt CLOUDY} software, with the goal of simultaneously reproducing UV and optical {\rm Fe{\sc ii}} emission.}
   {We employ a constant pressure model for the emitting clouds, instead of the customary constant density assumption. The allowed parameter range is broad, with metallicity up to 50 times the solar value and turbulent velocity up to 100 km~s$^{-1}$ for a subset of models. We also consider geometrical effects that could enhance the visibility of the non-illuminated faces of the clouds, as well as additional mechanical heating.}
   {Our investigation reveals that the broad line region of RM 102 is characterized by highly metallic gas. The observed {\rm Fe{\sc ii}} features provide strong evidence for an inflow pattern geometry that favours the dark sides of clouds over isotropic emission.}
   {This study confirms the presence of chemically enriched gas in the broad line region of bright quasars, represented by RM 102, which is necessary to explain the strong {\rm Fe{\sc ii}} emission and its characteristic features. Additionally, we report that {\tt CLOUDY} currently still lacks certain transitions in its atomic databases  which prevents it from fully reproducing some observed {\rm Fe{\sc ii}} features in quasar spectra.}


   \maketitle

\section{Introduction}

Singly ionized iron emission ({\rm Fe{\sc ii}}) is one of the most ubiquitous signatures in the spectra of Active Galactic Nuclei (AGN), with emission lines spanning from the ultraviolet (UV) to the near-infrared (NIR) wavelength ranges. Especially in the UV and optical ranges, the number of transitions associated with {\rm Fe{\sc ii}} is so high that they blend forming a pseudo-continuum. The presence of this pseudo-continuum significantly complicates the fitting of lines emitted from the Broad Line Region (BLR), creating a long-standing challenge \citep{sargent1968, collin1979, joly2008}. 

In the optical wavelength range, the intensity of {\rm Fe{\sc ii}} emission is connected to the critical $R_{\rm FeII}$ parameter of the Quasar Main Sequence (MS) \citep{Boroson92, sulentic2000}, which is defined as the ratio between {\rm Fe{\sc ii}} emission measured in the 4434-4684 \AA\ range and H$\beta$ emission. In the context of the MS, the sources that emit the largest amount of radiation per unit of mass are categorized as extreme-accreting sources, or extreme population A (xA) objects. These sources are characterized by $R_{\rm FeII}>1$ and a moderate-to-high Eddington ratio ($L/L_{\rm Edd}$), typically greater than 0.1 \citep{panda_etal_2019}, making them the strongest {\rm Fe{\sc ii}} emitters.

The spatial extent of the {\rm Fe{\sc ii}} emitting region remains a matter of debate. Some studies suggest that this emission originates in a region similar to that of H$\beta$ \citep{shapovalova2012, hu2015}, while others propose that the {\rm Fe{\sc ii}} emitting region may be more extended, potentially up to twice as large \citep{barth2013, panda_etal_2018, hu2020, he2021, lu2021}. However, the assumption of single-zone emission from the BLR seems appropriate, particularly for xA sources \citep{marziani2001, panda_etal_2019_QMS, panda2021}.

Historically, two main methods have been employed to measure {\rm Fe{\sc ii}} emission: (1) observational templates, created by isolating {\rm Fe{\sc ii}} emission from the rest of the known lines in the spectrum of a quasar \citep[e.g. I~Zw~1 for][]{vestergaard2001, marziani2009}; and (2) theoretical templates, which are based on calculating the transition rates of each emission line \citep{bruhweiler2008, panda_etal_2019_QMS}.

The challenge with observational templates is that they are limited by the signal-to-noise ratio (SNR) and the wavelength coverage of the spectra. Additionally, no existing template simultaneously covers both UV and optical bands \citep{Pandey2024, zhang2024}.
Theoretical templates, on the other hand, often suffer from inconsistencies between the measured emissions in the UV and optical ranges, frequently achieving a good fit in one range but showing strong mismatches in the other \citep{Pandey2024}.

In this work, we aim to obtain a working {\rm Fe{\sc ii}} template that fits the whole spectrum of bright quasar SDSS J141352.99+523444.2 from the UV to the optical bands. This object, hereafter referred to as RM 102, was selected from the SDSS Reverberation Mapping (SDSS-RM) catalogue \citep{Shen19, Pandey2024}.  RM 102 is an ideal candidate for this study because its high SNR spectrum spans a broad wavelength range (2000-5500 \AA), allowing for a comprehensive analysis of both the rest-frame UV and optical ranges. Furthermore, reliable estimates of the H$\beta$ and {\rm Mg{\sc ii}} time delays are available for this object, with $\tau_{{\rm H}\beta} = 104.6^{+20.5}_{-18.5}$ days and $\tau_{\rm MgII} = 101.7^{+11.6}_{-10.3}$ days \citep{shen2023}. There is strong evidence that {\rm Fe{\sc ii}} time delays are consistent with those of {\rm Mg{\sc ii}} and H$\beta$ \citep{zajacek24}. Moreover, its spectrum strongly resembles the quasar composite spectrum from \cite{vandenberk2001}, indicating that it is highly representative of the quasar population. 

In this study, we consider local processes which can affect the local {\rm Fe{\sc ii}} emissivity, such as radiative and mechanical heating, geometry (illuminated and dark sides of the clouds), microturbulence and metallicity, to better understand the physical conditions in typical quasars. 

We systematically analyze the {\rm Fe{\sc ii}} emission of RM 102, beginning with a description of  its SDSS spectrum in Section \ref{rm102}. Section \ref{methods} details the methodology and techniques used to disentangle the properties of our source. The results of our analysis are presented in Section \ref{results}. In Section \ref{discuss}, we discuss the main findings of our work. Finally, Section \ref{conclusions}, summarizes our conclusions and suggests directions for future research.

\section{Observed spectrum of RM 102}
\label{rm102}

\begin{figure*}[h]
    \centering
    \includegraphics[width=18.5cm]{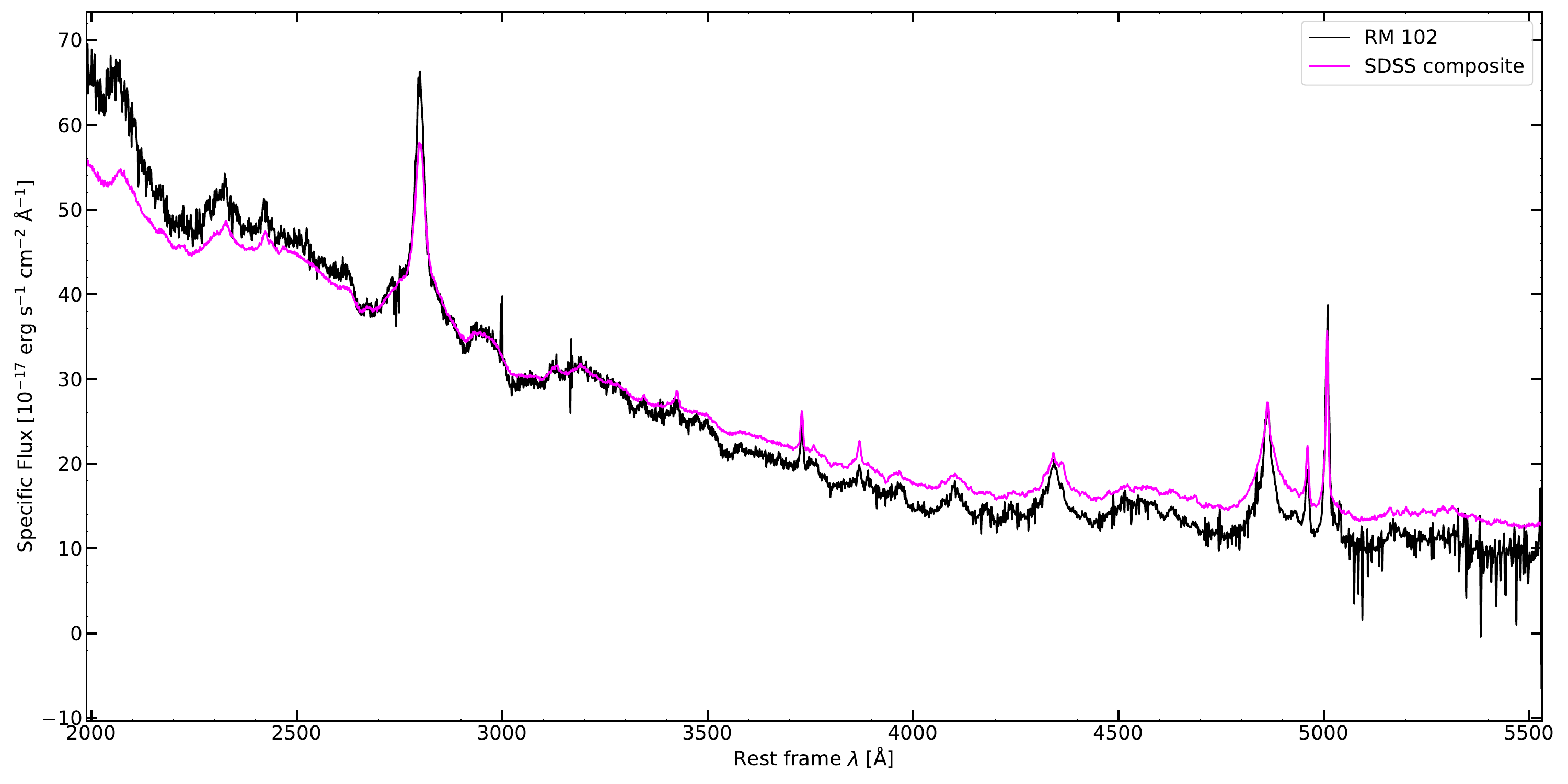}
    \caption{Plot of the observed spectrum of RM 102 (black), superimposed with the SDSS composite spectrum (magenta) from \cite{vandenberk2001}. The SDSS composite is normalized at $\lambda2800$ \AA.}
    \label{fig:fig1}
\end{figure*}

RM 102 was selected for this study due to its representative quasar characteristics, high-quality spectral data, and well-documented properties from reverberation mapping studies \citep{Shen19}. The spectrum used in this study is a compilation of 70 spectra from the SDSS Science Archive Server (SAS), the DR16~\footnote{\url{https://dr16.sdss.org/optical/spectrum/search}}. The source did not exhibit significant variability in spectral shape, justifying the use of a compiled median spectrum to achieve a high SNR. The resulting spectrum is shown in Figure \ref{fig:fig1}, with the SDSS composite spectrum from \citet{vandenberk2001} overlaid for comparison. The two spectra show a remarkable similarity in both the overall continuum shape and line properties, although RM 102 exhibits stronger {\rm Fe{\sc ii}} emission than the average quasar in the \citet{vandenberk2001} composite spectrum. This suggests that RM 102 is highly representative of bright quasars, particularly those in the xA category.

The global properties of RM 102 are summarized in Table~\ref{tab:RM102_info}, which lists its redshift ($z$), bolometric luminosity ($L_{\rm bol}$), black hole mass ($M_{\rm BH}$), Eddington ratio, $R_{\rm FeII}$ parameter, the measured time delay to H$\beta$ ($\tau_{{\rm H}\beta}$) and the estimated ionizing flux $\Phi$(H) from the central engine. The Eddington ratio is slightly higher than the average value of 0.17 in the SDSS quasars \citep{panda_etal_2018}. Moreover, the $R_{\rm FeII}=1.24$ classifies RM 102 as a "bona fide" xA source, as its $R_{\rm FeII}$ exceeds the lower limit of 1.0 associated with extreme Population A sources in the context of the MS \citep{sulentic2000, duetal16a, marziani2018, panda_etal_2019_QMS}.

The same spectrum compiled spectrum was previously used in preliminary tests of the {\rm Fe{\sc ii}} templates by \citet{Pandey2024}. In this work, we aim to conduct a more thorough analysis of both the broad-band properties and the individual spectral features of RM 102.

\begin{table}[h]
\centering
\caption{Basic properties of the RM 102.}
\begin{tabular}{l  c  r}
\hline\hline
Parameter & Value & Reference \\
(1) & (2) & (3) \\
\hline
$z$ & 0.86114$\pm$0.0003 & S19\\
$\log L_{\rm bol}$ [erg s$^{-1}$] & 45.7179$\pm$0.0005 & S19\\  
$\log M_{\rm BH}$ [$M_\odot$] & 8.086$\pm$0.036 & S19 \\
$L/L_{\rm Edd}$ & 0.329$\pm$0.027 & S19\\
$R_{\rm FeII}$ & 1.24$\pm$0.70 & P24\\
$\tau_{{\rm H}\beta}$ [days] & $104.6^{+20.5}_{-18.5}$ & S23\\
$\log\Phi$(H) [cm$^{-2}$~s$^{-1}$] & $20.43$ & P24\\
\hline
\end{tabular}
\tablefoot{(1) Parameters defining the basic properties of RM 102. (2) Value of the parameter. (3) Paper from which the reference is obtained, with S19 referring to \cite{Shen19}, P24 referring to \cite{Pandey2024}, and S23 referring to \citet{shen2023}.}
\label{tab:RM102_info}
\end{table}




\section{Methodology}
\label{methods}

\subsection{CLOUDY photoionization simulations}
\label{sect:CLOUDY_sim}

In this study, we build upon the methodology employed by \citet{Pandey2024} for grid creation and analysis of the same quasar, RM 102. However, considering that \citet{Pandey2024} did not achieve a satisfactory explanation of the broadband {\rm Fe{\sc ii}} spectrum, we have significantly refined and modified our approach to better understand the emission properties of the source.

Similar to \citet{Pandey2024} we utilize the most recent version of the spectral synthesis code {\tt CLOUDY} C23.00 \citep{Cloudy23}, which features the most extensive database of {\rm Fe{\sc ii}} transitions available to date. Motivated by the arguments of cloud stability, formation mechanisms and pressure confinement as discussed in various studies \citep[e.g.][]{begelman1990,goncalves2006,czerny2009,stern2014,adhikari2018,borkar2021}, we assume constant pressure instead of constant density for the clouds.

In a simplified model, a BLR cloud can achieve pressure equilibrium, where its internal thermal pressure is balanced by external pressures. If this equilibrium coexists with balanced heating and cooling processes, the cloud can also be in thermal equilibrium. However, in more complex and realistic scenarios, achieving simultaneous pressure and thermal equilibrium is challenging due to varying radiation fields and gradients in temperature, density, or ionization states. Thus, while a cloud in pressure equilibrium may not be in perfect thermal equilibrium, both conditions can coexist to a reasonable approximation in localized regions of the cloud. To explore these dynamics, we conduct simulations under both constant pressure and constant density assumptions to compare outcomes.


{\tt CLOUDY} allows for the simulation of different temperature and density conditions on both the illuminated and dark sides of a cloud independently, making it a suitable tool for our study. The total pressure $P_{\rm tot}$ for each cloud is computed as: 

\begin{equation}
    P_{\rm tot} = P_{\rm gas} + P_{\rm turb} + P_{\rm lines} + \Delta P_{\rm rad},
\end{equation}
where $P_{\rm gas}$ is the thermal gas pressure, $P_{\rm turb}$ is the turbulent pressure and $P_{\rm lines}$ is the radiation pressure due to trapped emission lines \citep{ferland1984}. The turbulent pressure is calculated as $P_{\rm turb}=\rho V_{\rm turb}^2/2$, where $\rho$ is the gas density and $V_{\rm turb}$ is the microturbulent velocity. The radiative pressure difference, $\Delta P_{\rm rad}$, due to the attenuated incident continuum, is calculated as $\Delta P_{\rm rad} = \int a_{\rm rad} \rho dr$, with $a_{\rm rad}$ being the radiative acceleration.

We explore a grid of simulations with $\log\Phi$(H) ranging from 17 to 22 cm$^{-2}$~s$^{-1}$, and with pressure covering the $\log (P_{\rm tot})$ from 14 to 18 cm$^{-3}$ K range, adopting a step size of 0.25 dex for both parameters. The values of $V_{\rm turb}$ are varied from 0 to 40 km~s$^{-1}$ with a step-size of 10 km~s$^{-1}$. Models with a microturbulence value of 100 km~s$^{-1}$ were also studied. The metallicity $Z$ ranges from 1 to 50 times solar metallicity (Z$_{\odot}$) in steps of 1, 2, 5, 10, 20, and 50 Z$_{\odot}$, to cover a wide range of potential chemical enrichment scenarios.

\subsection{CLOUDY mechanical heating}
\label{mech}

In addition to heating from the central source via ionizing flux $\Phi$(H), we explore the possibility that a portion of the heating within the BLR clouds arises from mechanical processes, specifically collisions between clouds. This scenario assumes that mechanical energy is injected into the system as clouds with relative motion collide. We model these relative motions as microturbulence, which could represent a small fraction of the Keplerian orbital velocity of the clouds around the supermassive black hole (SMBH).

Our order-of-magnitude evaluation of this effect assumes that the duration of the collision is roughly given by the sound speed crossing time across the cloud. 
From the {\tt CLOUDY} inputs we can estimate the size of a BLR cloud as $R_{\rm cl} = N_{\rm c}/n_{\rm H}$, where $R_{\rm cl}$ is the cloud radius, $N_{\rm c}$ is the column density and $n_{\rm H}$ is the Hydrogen density. The collision timescale ($t_{\rm col}$) for two clouds is thus computed as:

\begin{equation}
    \centering
    t_{\rm col} = R_{\rm cl}/V_{\rm turb},
\end{equation}
and the mass of each cloud is: 
\begin{equation}
    \centering
    M_{\rm cl} = \frac{4}{3}\pi \rho R_{\rm cl}^3,
\end{equation}
where $\rho$ is the cloud density, assuming all gas is composed of Hydrogen.
The heating $H$ is thus calculated as 
\begin{equation}
    \centering
    H = \alpha \frac{M_{\rm cl} V_{\rm turb}^2}{R_{\rm cl}^3 t_{\rm col}},
\end{equation}
where $\alpha$ is an arbitrary parameter.
The adopted dimensionless parameter scaling this quantity is generally much smaller than 1, as the clouds collide frequently but not constantly. Some of the clouds may collide not only with each other but also with the disk \citep[e.g.][]{muller2022}.
We assume a constant heating throughout the cloud volume.

Assuming the $\alpha$ parameter to be 0.01, a conventional BLR cloud density of $n_{\rm H}=10^{11}$ cm$^{-3}$ and column density $N_c= 10^{24}$ cm$^{-2}$, the additional mechanical heating resulted from the equations in Section \ref{mech} is $H=10^{-7}$ erg cm$^{-3}$ s$^{-1}$, which will be employed in this work when the addition of extra mechanical heating is considered.

\subsection{Comparison with the observational data}
\label{sect:fitting}

To compare our models with observational data, we first corrected the fluxes of RM 102 for galactic extinction using the extinction law described by \cite{cardelli1989}. We adopted the absorption value in the $V$ band of $A_V = 0.026$ mag, as provided by the NED extragalactic database.  The corrected spectrum was then fitted following the methodology outlined in \citet{Pandey2024}. This process involved decomposing the spectrum into its underlying power-law continuum, emission lines, and {\rm Fe{\sc ii}} contributions. 

In our analysis, we modeled the broad components of lines using a Lorentzian profile and included a semi-broad component for the [OIII]$\lambda \lambda4959,5007$ doublet. The presence of a strong Balmer continuum was not requested as such a component is not clearly seen in the data. This omission aligns with previous studies, which show that AGNs exhibit a range of Balmer edge strengths. The strength of the Balmer continuum varies significantly, with the feature's average strength (as measured at 3000 \AA) around 0.1 relative to the flux before after Balmer continuum subtraction, and with a range from 0 to 0.18 among a sample of 287 objects \citep{kovacevic2017}.

To avoid underestimating the emission line contributions, we used semi-empirical UV/optical {\rm Fe{\sc ii}} templates implemented in the fitting code PyQSOFit \citep{pyqsofit}. We modified the approach by dividing the wavelength range into eight segments, allowing the {\rm Fe{\sc ii}} flux to vary individually in each segment, adopting different normalization factors. The error in normalization varies from 25 \% to nearly 70 \% across the segments (see Table \ref{tab:RM102_errors}) . The slope of the power-law continuum was fixed, and the best-fit slope value was 2.08. This increased the fit errors but, within the error, accounted for possible departures from a single power law in the spectrum. Such departures could result from starlight contamination at longer wavelengths or, at shorter wavelengths, from internal reddening or proximity to the accretion disk temperature peak. We will discuss these issues in Section \ref{undcont}. The modification of the {\rm Fe{\sc ii}} fitting methodology led to a reduction in the reduced $\chi^2$ value of 8\% with respect to \citet{Pandey2024} results (see their Fig.~14). The level at 2800 \AA~ is now higher, not dropping to zero, although the UV emission still remains significantly stronger than the optical emission.

During the fitting, the Full Width at Half Maximum (FWHM) of lines was determined, and the measured FWHM of H$\beta$ was found to be approximately 1600 km s$^{-1}$, consistent with an xA source in the context of the MS \citep{Boroson92, sulentic2000}. The FWHMs of {\rm Fe{\sc ii}} lines are similar, evidencing that the two emissions are likely produced at similar distances from the SMBH.

To select models most suitable for representing the {\rm Fe{\sc ii}} emission in RM 102, we employed several approaches. First, we used the broadband integrated {\rm Fe{\sc ii}} emissivity, selecting four bands as in \citet{Pandey2024}: two in the UV band (UV red and blue wings), on either side of the {\rm Mg{\sc ii}} line, and two in the optical band (optical red and blue wings), on either side of the H$\beta$ line. We measured the integrated fluxes from the data, as described in more detail in \citet{Pandey2024}.\\

\begin{table}[h]
\centering
\caption{Template ranges and errors.}
\begin{tabular}{c c c}
\hline\hline
Template & Spectral range & Normalization factor \\
 & [\AA] &\\
(1) & (2) & (3)  \\
\hline
a,b & 2000-2650 & $15.57\pm5.73$ \\
b & 2650-2800 & $13.52\pm8.83$ \\
b & 2800-3090 & $17.83\pm7.22$ \\
c & 3090-3500 & $16.78\pm3.88$ \\
d & 3686-4434 & $2.68\pm1.88$ \\
d & 4434-4684 & $6.03\pm2.36$ \\
d & 4684-5150 & $6.94\pm2.89$ \\
d & 5150-5350 & $7.14\pm2.72$ \\
\hline
\end{tabular}
\tablefoot{(1) Template used for the fitting in the identified spectral range: a: \citet{vestergaard2001}, b:\citet{salvainder06}, c:\citet{Tsuzuki2006}, d: \citet{Boroson92} . (2) Wavelength range associated with the template. (3) Best fitting normalization factor for the selected template, with error.}
\label{tab:RM102_errors}
\end{table}

Specifically, we considered the following wavelength ranges of primary importance:

\begin{enumerate}
    \item $2000$-$3090$ \AA\ and $4000$-$5350$ \AA\ for the UV and optical measurements, respectively. The UV/optical ratio observed for RM 102 is $1.897\pm0.645$.

    \item $2800$-$3090$ \AA\ and $2650$-$2800$ \AA\ for the red-to-blue wings ratio of {\rm Fe{\sc ii}} in the UV, respectively. The red/blue UV ratio observed for RM 102 is $2.178\pm1.677$. 
    
    \item $5150$-$5350$ \AA\ and $4434$-$4684$ \AA\ for the red-to-blue wings ratio of {\rm Fe{\sc ii}} in the UV, respectively. The red/blue optical ratio observed for RM 102 is $0.734\pm0.404$. 
\end{enumerate}

The theoretically calculated spectra allow us to determine the specific ratios, $R$, between two bands. Since we calculate both the inward and outward emission, the specific geometrical setup allows for relative enhancement, so we introduce a relative factor, $A$, between the two. In the standard geometry, when both the illuminated and shielded faces of the clouds equally contribute, $A=\frac{1}{2}$. In addition, we can combine $N_{cl}$ clouds with different properties (e.g. ionizing flux, density, turbulent velocity, external heating etc.). The most general expression for the ratio is thus:
\begin{equation}
R = \frac{\sum_{n=1}^{N_{cl}} ((1-A) F_{k,band1}^{inward} + A F_{k,band1}^{outward})} {\sum_{n=1}^{N_{cl}}((1-A) F_{k,band2}^{inward} + A F_{k,band2}^{outward})},
\end{equation}
where $F_{band1}^{inward}$ is the integrated inward flux in the band1 for cloud $k$, and $F_{band1}^{outward}$ is the integrated outward flux for cloud $k$.

We compared these ratios with those derived from other {\rm Fe{\sc ii}} templates available in the literature. For example, the UV red-to-blue ratio in the UV templates of \citet{popovic2019} without any rescaling of the components, is 2.8. This value is slightly higher than our measurement for RM 102, but remains consistent within the margin of error. Similarly, the optical template based on I~Zw~1 \citep{marziani2009} gives a red-to-blue ratio of 0.772, which aligns well with our measurement for RM 102.

We can use a single such ratio, or a linear combination of $i$ such ratios to compare the model to the data quantitatively, using the $\chi^2$ statistic.

The $\chi^2$ is computed as 
\begin{equation}
\label{chi}
    \chi^2 = \sum_{\rm i} \left(\frac{R_{\rm i}-R_{\rm i,mod}}{\delta R_{\rm i}}\right)^2,
\end{equation}
with $i$ representing different ratios. Here, $\delta R_{\rm i}$ is the measurement error of the ratio from the observed spectrum. Most of the time we use a single cloud representation ($N_{cl} = 1$), and equal contribution of inward and outward radiation ($A=\frac{1}{2}$).

\section{Results}
\label{results}

\subsection{Properties of the constant pressure clouds without and with mechanical heating}
\label{sect:constant_pressure}

\begin{figure*}[h]
    \centering
    \includegraphics[width=18.5cm]{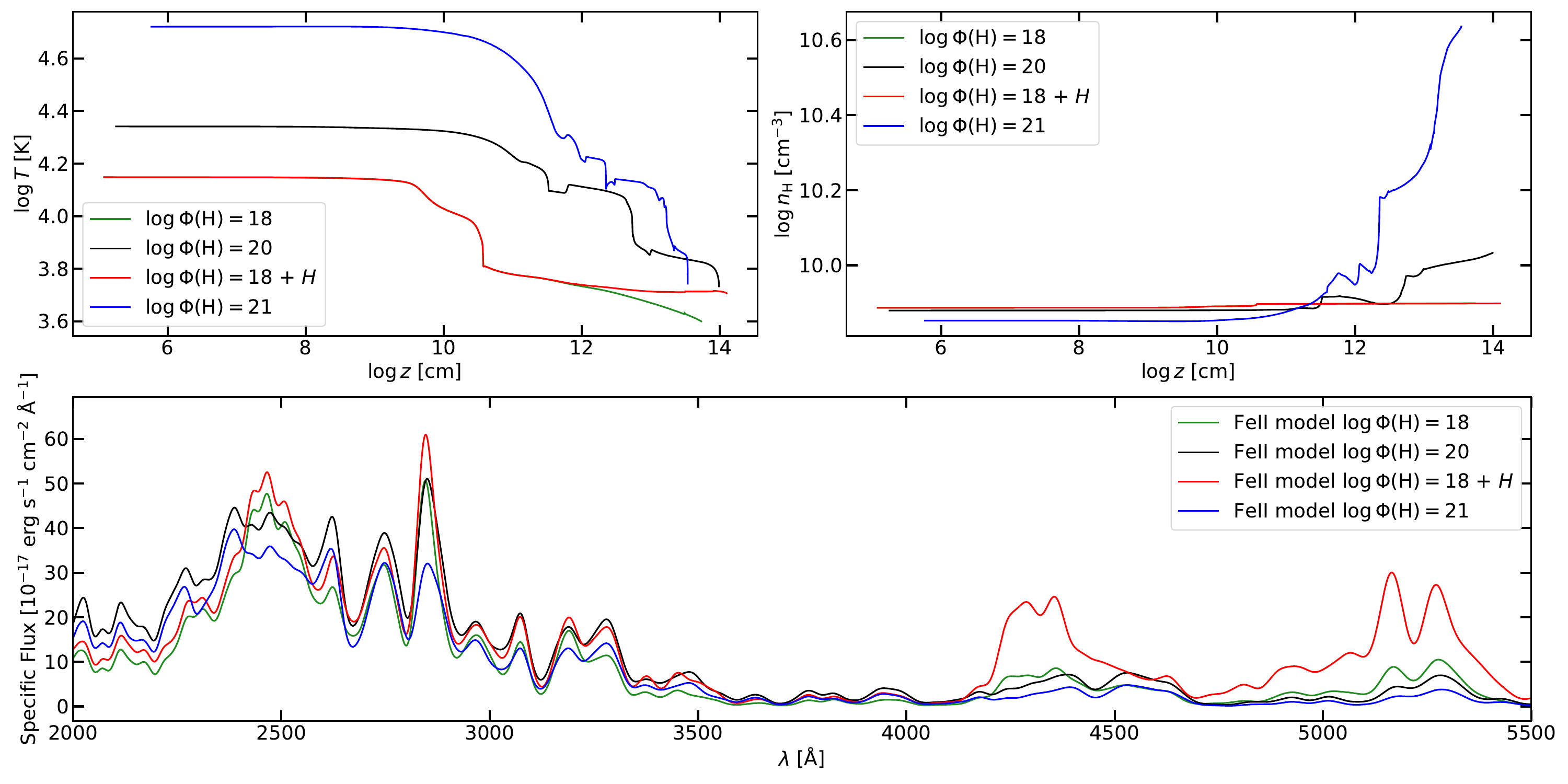}
    \caption{Structure and spectra of exemplary clouds simulated with {\tt CLOUDY} at constant pressure, assuming $\log (P_{\rm tot})=16$ cm$^{-3}$ K, $V_{turb} = 100$ km~s$^{-1}$, $N_c=10^{24}$ cm$^{-2}$ and $Z=10Z_\odot$. The assumed values of ionizing fluxes are $\log\Phi$(H)=18 cm$^{-2}$~s$^{-1}$ (displayed in green), 20 cm$^{-2}$~s$^{-1}$ (black) and 21 cm$^{-2}$~s$^{-1}$ (blue). An additional example with $\log\Phi$(H)=18 cm$^{-2}$~s$^{-1}$ and the addition of extra mechanical heating is displayed in red. {\it Top left}: Cloud temperature profile in logarithmic scale as a function of depth ($z$) of the cloud. {\it Top right}: Cloud density profile in logarithmic scale as a function of depth ($z$) of the cloud. {\it Bottom}: Spectra of {\rm Fe{\sc ii}} emission from different models, displayed with 100\% covering factor. Each synthetic spectrum is broadened with a Gaussian kernel with FWHM = 1600 km s$^{-1}$.}
    \label{fig:str}
\end{figure*}

Most {\tt CLOUDY} simulations and {\rm Fe{\sc ii}} templates utilize the constant density approximation for a single cloud. To illustrate the properties of clouds under conditions of constant pressure, Figure \ref{fig:str} presents examples of the temperature and density structure across the cloud. One example represents a highly ionized cloud with $\log\Phi$(H) = 20 cm$^{-2}$~s$^{-1}$, and another one represents a weakly ionized cloud with $\log\Phi$(H) = 18 cm$^{-2}$~s$^{-1}$. The corresponding spectra, assuming equal contributions from both the dark and bright sides, are shown in the lower panel of Figure \ref{fig:str}.

For highly ionized clouds, the temperature drop across the cloud from the illuminated part towards its back is quite steep, accompanied by a rapid density increase, although the overall density difference between the two sides is only about 40\%. In contrast, weakly ionized clouds exhibit a significant temperature drop across the cloud, while the density remains nearly constant throughout. These variations also affect the inward and outward spectra. For highly ionized clouds, the two spectra are relatively similar. However, for weakly ionized clouds, the outward radiation is much weaker, with the most substantial difference occurring in the UV part of the spectrum, which is significantly fainter compared to the illuminated side of the cloud.

When examining an even stronger ionizing flux, $\log\Phi$(H) = 21 cm$^{-2}$~s$^{-1}$, we do not observe a simple monotonic behaviour relative to the previous cases, and the temperature and density gradients across the cloud are steeper. Notably, at this higher ionization level, the physical conditions become less favorable for {\rm Fe{\sc ii}} emission, resulting in a flux weaker than that seen in the $\log\Phi$(H) = 20 cm$^{-2}$~s$^{-1}$ case. For certain transitions in the UV and optical ranges, the flux is even lower than for the $\log \Phi$(H) = 18 cm$^{-2}$~s$^{-1}$ case.

When the predicted mechanical heating from the collision of clouds is included ($H= 10^{-7}$ erg cm$^{-3}$ s$^{-1}$), the overall ionization level increases and the temperature gradient becomes shallower. However, this change is only evident in clouds exposed to weaker ionizing flux. The change in the temperature profile is only noticeable deep within the cloud, but on a linear scale, this would represent most of the cloud's body. Clouds exposed to stronger ionizing flux ($\log\Phi$(H) $\geq 20$ cm$^{-2}$~s$^{-1}$) are unaffected by the presence of extra mechanical heating, as the impact of this heating is likely negligible. Consequently, the spectrum and the temperature and density profiles of the clouds with $\log\Phi$(H) = 20-21 cm$^{-2}$ s$^{-1}$ are not displayed in Figure \ref{fig:str}, since they remain unchanged compared to the cases without mechanical heating. 

In contrast, for weakly ionized clouds, the presence of mechanical heating increases the temperature of the cloud's dark side relative to clouds without additional heating. This results in a noticeable increase in the emitted flux in the optical part of the spectrum, and consequently a potential reduction in the UV-to-optical ratio. It is evident that the dark sides of clouds significantly contribute to the shape of the optical {\rm Fe{\sc ii}} emission. This result is in agreement with previous studies suggesting that very low ionization lines, such as those of optical {\rm Fe{\sc ii}}, originate from shielded regions \citep{joly87, ferland89}. Meanwhile, the interpretation of the UV spectrum is more complex due to the larger number of transitions and mechanisms involved \citep{wills1985}. 

\subsection{The role of turbulent velocity in constant density and constant pressure approaches in simulations}
\label{turbdp}

\begin{figure*}[h]
    \centering
    \includegraphics[width=18.5cm]{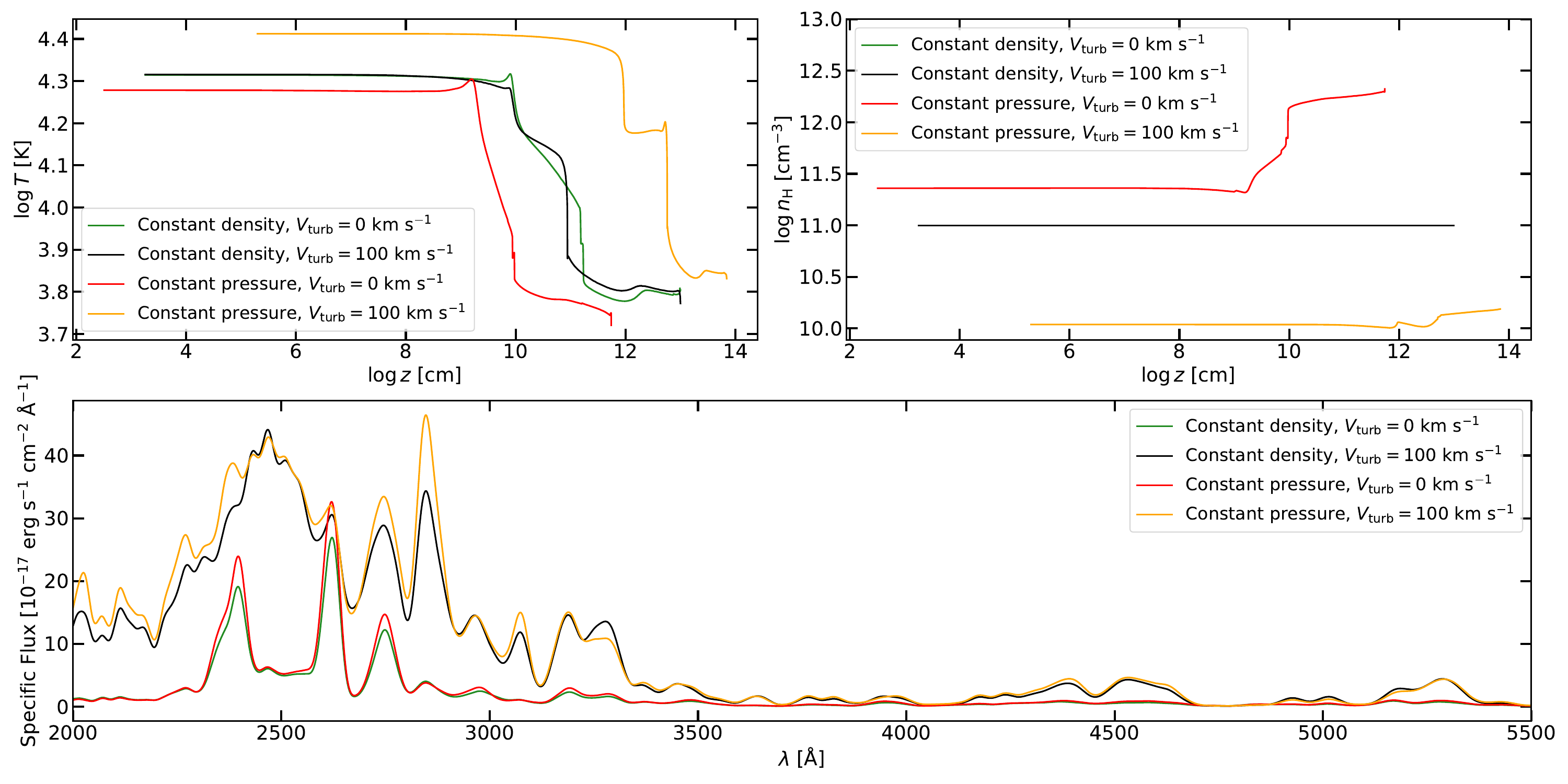}
    \caption{Structure and spectra of exemplary clouds simulated with {\tt CLOUDY} at constant density (green and black), and constant pressure (red and orange), with different values of $V_{\rm turb}$, assuming similar physical parameters. The ionizing flux was assumed $\log \Phi$(H)=20 cm$^{-2}$~s$^{-1}$ for all clouds, and metallicity is assumed solar. Clouds in conditions of constant density have density fixed at $\log n_{\rm H}= 11$ cm$^{-3}$, while in case of constant pressure $\log (P_{\rm tot}) =16$ cm$^{-3}$ K. {\it Top left}: Cloud temperature profile in logarithmic scale as a function of depth ($z$) of the cloud. {\it Top right}: Cloud density profile in logarithmic scale as a function of depth ($z$) of the cloud. {\it Bottom}: Spectra of {\rm Fe{\sc ii}} emission from different models, displayed with 100\% covering factor. Each synthetic spectrum is broadened with a Gaussian kernel with FWHM = 1600 km s$^{-1}$.}
    \label{fig:str_constants}
\end{figure*}

Continuing with the assumption of constant pressure as detailed in Section~\ref{sect:CLOUDY_sim}, we compare clouds simulated with these conditions with those simulated under the assumption of constant density. Figure \ref{fig:str_constants} highlights the differences between these various simulations with similar physical conditions typically associated with the BLR. Two cases assume constant density, while the other two assume constant pressure. Additionally, we examine the impact of high turbulent velocity, $V_{\rm turb} = 100$ km~s$^{-1}$, in one case for constant density and one for constant pressure.

In all scenarios, the clouds exhibit a complex temperature structure. However, the density is either constant or determined by the {\tt CLOUDY} code under the condition of pressure equilibrium. This characteristic is directly linked to the role of turbulent velocity within the cloud. In the constant density scenario, for a given column density, the cloud size remains unchanged regardless of the turbulent velocity. In this case, increased turbulent velocity primarily enhances the gas emissivity without significantly altering the temperature or density profiles. In contrast, under constant pressure, turbulent velocity influences the temperature profile, which subsequently affects the density profile. As turbulent velocity increases, the density decreases and the cloud size expands for a fixed column density. Specifically, the diameter of the cloud increases by about two orders of magnitude in the presence of turbulent velocity.

The resulting spectra from these simulations are shown in the bottom panel of Figure \ref{fig:str_constants}. These spectra reveal that turbulent velocity plays a crucial role in shaping the {\rm Fe{\sc ii}} emission. Notably, optical {\rm Fe{\sc ii}} emission is not significantly produced in the absence of turbulent velocity under both constant density and constant pressure conditions. On the other hand, the conditions of constant density or constant pressure do not lead to significant variations in the shape of the emitted {\rm Fe{\sc ii}}.

\subsection{Fits of the three broad-band flux ratios without mechanical heating}
\label{sect:three_ratios}

\begin{figure*}[h]
    \centering
    \includegraphics[width=18.5cm]{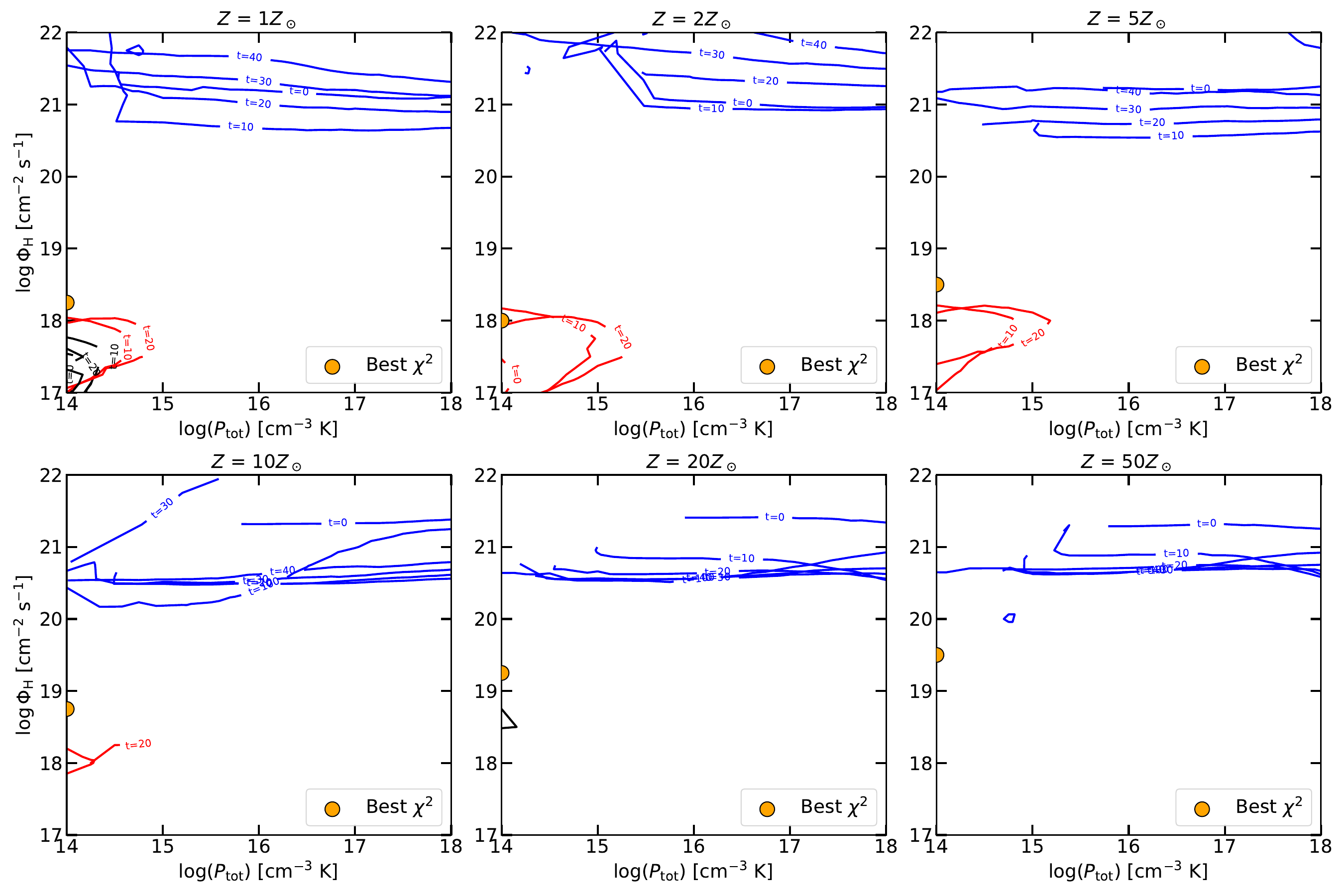}
    \caption{Contour plots denoting the ratios observed in the source RM 102 for different values of metallicity and microturbulence. The black, red, and blue contours denote the UV-to-optical {\rm Fe{\sc ii}} ratio, UV red-to-blue {\rm Fe{\sc ii}} ratio, and optical red-to-blue {\rm Fe{\sc ii}} ratio, respectively. An orange dot in each plot marks the position of the best $\chi^2$ for each bin of metallicity.}
    \label{fig:turbplots5050}
\end{figure*}

We begin by analyzing the first three global ratios (1) to (3) described in Section~\ref{sect:fitting}. 

Using the computed grid and the measured fluxes for RM 102, we identify the parameters (density and ionization flux) which best reproduce the observed value for each of the three ratios. The contour plots for each ratio are shown and discussed in Appendix~\ref{app:3ratios}. The different panels do not always show the lines corresponding to the measured ratio, but the errors on the ratios are large (see Section~\ref{sect:fitting}), resulting in a broad allowed parameter space for each ratio. The best fit for the data, considering each ratio separately, typically corresponds to different simulation parameters.

We summarize these findings by showing the plots for these three ratios in Figure~\ref{fig:turbplots5050}. The lack of intersection between the red and blue lines indicates that a single density and ionization flux combination does not reproduce both the UV red-to-blue ratio and the optical red-to-blue ratio for RM 102. The observed overall UV-to-optical ratio only falls within the adopted grid when the density (or pressure) is very low. The range expands slightly when metallicity is increased to 20 times the solar value, but even then, the UV red-to-blue ratio is not reproduced. We can still identify the best solution (orange dot) in each of the contour plots which minimizes the $\chi^2$ fit for all these ratios. The lowest value of the $\chi^2$ in Figure~\ref{fig:turbplots5050} is 3.81, for $20Z_\odot$. 

We select the best solution for each metallicity value, and the best of all corresponds to the $Z=20 Z_\odot$, $V_{turb} = 40$ km~s$^{-1}$, $\log\Phi$(H) = 19.25 cm$^{-2}$~s$^{-1}$ and $\log(P_{\rm tot}) = 14$ cm$^{-3}$ K (see first line in Table \ref{tab:solutions}). For this solution we plot the modelled {\rm Fe{\sc ii}} emission against the observed fluxes in RM 102 (see Figure~\ref{fig:best3}). Using the observed flux, we still have the freedom to assume an arbitrary covering factor $f$ to match the data. We selected 0.24, as this value of $f$, for this solution,  matches the integrated intensity of {\rm Fe{\sc ii}} emission of RM 102 in the selected wavelength range. Such covering factor is well within the expected range of the BLR covering factors, 0.1 - 0.3 (e.g. \citealt{peterson_review2006,collin2006}; see also \citealt{naddaf_CF2024} for model-based estimates). Despite this formal acceptance, the actual representation of the {\rm Fe{\sc ii}} emission is far from perfect. The UV data are marginally matched, but the optical emission is underproduced, particularly in the blue wing of the optical {\rm Fe{\sc ii}}. Moreover, a strong emission feature that is absent in the observed spectrum is present at about 4300\AA. Other discrepancies include the lack of a dip in emission at approximately 2400\AA, and a significant underproduction of {\rm Fe{\sc ii}} in the 3000-3500\AA\ range. The model predicts two clear spikes at approximately 2350 \AA~ and $\sim 2610$ \AA, separated by a gap. This feature has been studied in previous works \citep{baldwin2004,sarkar2021}, and is discussed more in detail in Section \ref{transitions}. In the far UV, the same transitions appear in both the model and the data but with different intensities, suggesting that the physical conditions represented by the best-fit model do not yet fully reproduce the entire ionization balance in the medium.

When calculating the minimum $\chi^2$ for each metallicity bin, we notice that only the minimum $\chi^2$ associated with $Z=2Z_\odot$ is departing significantly from the minimum $\chi^2$ solution observed at $Z=20Z_\odot$. Indeed, only solutions with $5Z_\odot \leq Z \leq 20Z_\odot$ seem appropriate in the context of xA quasars \citep{panda_etal_2019_QMS, panda_etal_2020, floris2024}. Despite the large errors in the local subtraction of the underlying power-law, the fit is formally acceptable.

The overall consistency of the UV and optical emission in some specific ranges, assuming the predicted covering factor, is nonetheless notable given the adopted technique. However, the favoured ionization parameter $\log\Phi$(H) = 19.25 cm$^{-2}$~s$^{-1}$, is inconsistent with the observed SED of the object and the measured time delay of H$\beta$, as discussed in \citet{Pandey2024}. Such high values of ionizing flux would require substantial internal shielding of the radiation emitted from the central engine, implying that the radiation received by the BLR clouds differs from the measured continuum emission. Under the favored solution, the cloud density ranges from $3.27\times 10^7$ cm$^{-3}$ at the illuminated face to $3.60\times 10^{8}$ cm$^{-3}$ at the dark side. Such low densities are not typically associated with BLR conditions, suggesting that this solution may not be fully appropriate.

\begin{figure*}[h]
    \centering
    \includegraphics[width=18.5cm]{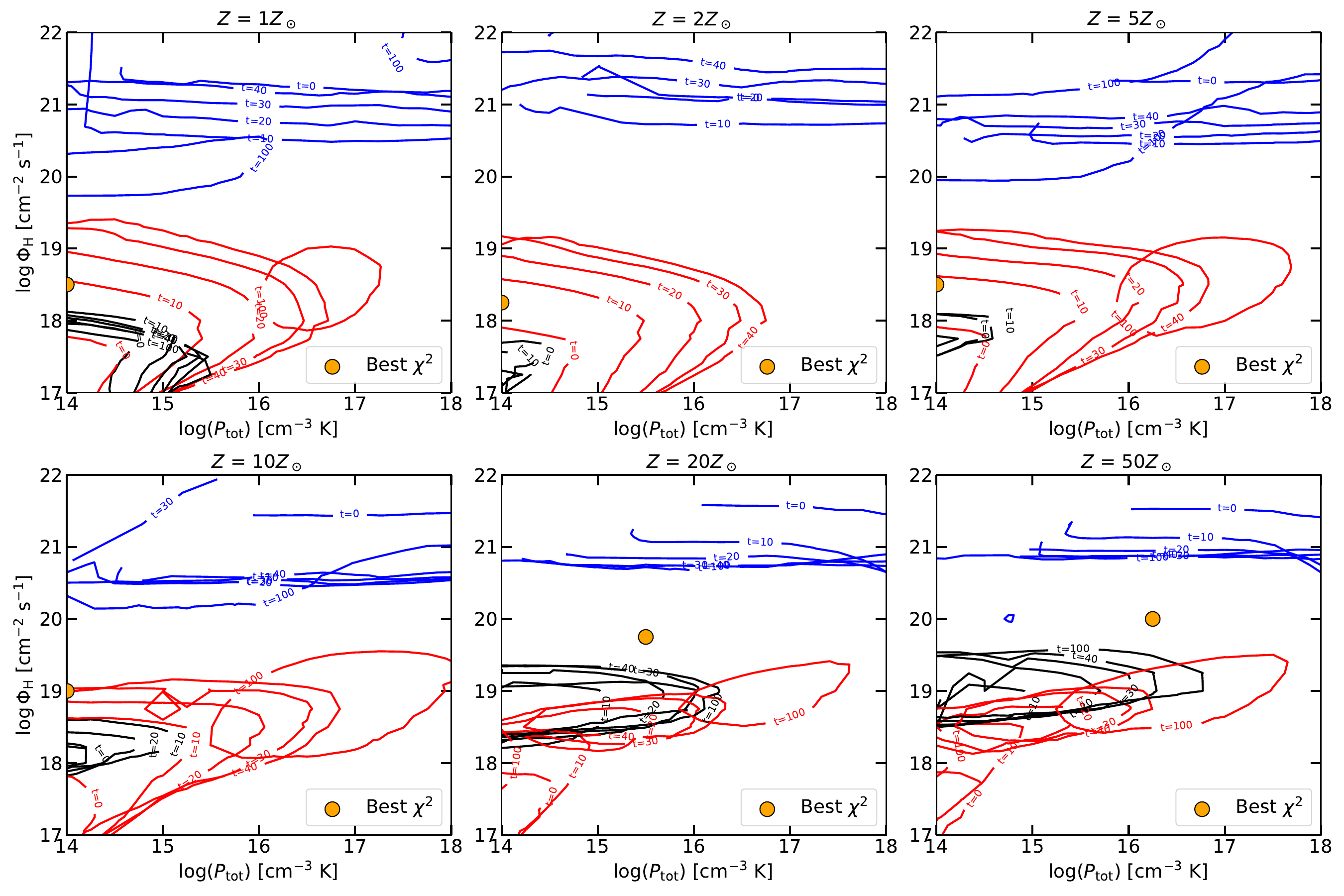}
    \caption{Contour plots denoting the ratios observed in the source RM 102 for different values of metallicity and microturbulence, assuming 20\% contribution from the bright side of the cloud and 80\% from the dark side. The {\rm Fe{\sc ii}} UV-to-optical ratio contours are shown in black, the {\rm Fe{\sc ii}} UV red-to-blue ratio contours are shown in red, and the {\rm Fe{\sc ii}} optical red-to-blue ratio contours are shown in blue. The position associated with the best $\chi^2$ for each bin of metallicity is marked with an orange dot.}
    \label{fig:turbplots2080}
\end{figure*}

We also constructed similar contour plots (see Figure \ref{fig:turbplots2080}) for the case when 20\% of the light came from the illuminated sides of the clouds and 80\% from the dark sides, corresponding to $A = 0.8$. This configuration reflects a geometry where some clouds are obscured, such as in an inflow pattern \citep{horne2021}. The new contour plots (see Figure~\ref{fig:turbplots2080}) reveal some changes compared to the previous case, with the best fit found for a slightly different set of parameters (see Table \ref{tab:solutions}), and a significantly reduced $\chi^2$ of 1.98. 

The model for the broad band is compared to the data in Figure~\ref{fig:best3} in Appendix \ref{app:minchifig}. Although the UV {\rm Fe{\sc ii}} emission nearly matches, some features in the UV region are not well represented, and the optical blue wing is strongly underproduced in the spectrum obtained from the {\tt CLOUDY} simulations. Despite these issues, the improvement in $\chi^2$ compared to the previous case with equal contributions from both cloud sides is consistent with the enhancement in the overall shape of the template, even if it still does not fully represent the data. The physical conditions associated with this {\tt CLOUDY} simulation appear more realistic than in the previous case, requiring less intense shielding from the central engine. Additionally, the cloud density in this scenario ($n_{\rm H}\sim10^9$ cm$^{-3}$) is still relatively low. While such low densities were once considered representative of the BLR, later studies suggested lower densities for high ionization lines and higher densities for low ionization lines like H$\beta$, {\rm Mg{\sc ii}} and {\rm Fe{\sc ii}}, starting from the end of the 1980s \citep{collin1988,collin_lasota1988}. Most of the recent works points towards much higher densities in the bulk of BLR \citep[e.g.][]{panda_etal_2019_QMS}. 

Low density values for the BLR clouds of RM 102 could be confirmed if a strong emission of {\rm C{\sc{iii]}}}$\lambda1909$ were detected for the object, as {\rm C{\sc{iii]}}} has a critical density of $n_{\rm c}$$\sim$ 10$^{10}$ cm$^{-3}$ \citep{hamann2002}. This line is clearly present in the composite quasar spectrum of \citet{vandenberk2001}. Unfortunately, the SDSS spectrum of RM102 used in this work ends at $\sim 2000$\AA. In addition, the {\rm C{\sc{iii]}}} line might originate closer in than {\rm Fe{\sc ii}}, as it is considered an intermediate line between low and high ionization lines \citep[see e.g.][]{buendia2023}.

\begin{table*}[h]
\centering
\caption{Best fitting solutions}
\begin{tabular}{l c c c c c c c c r}
\hline\hline
Ratios &Side contributions & $\log\Phi$(H) & $\log P_{\rm tot}$ & $V_{\rm turb}$ & $Z$ & $\log H$ & $f_{\chi^2}$ & $\chi^2_{\rm min}$ & $f$\\
& & [cm$^{-2}$ s$^{-1}$] & [cm$^{-3}$ K] & [km s$^{-1}$] & [$Z_\odot$] & [erg cm$^{-3}$ s$^{-1}$] & & & \\
(1) & (2) & (3) & (4) & (5) & (6) & (7) & (8) & (9) & (10)\\
\hline
Three & 50-50 & 19.25 & 14 & 40 & 20 & -- & -- & 3.81 & 0.24\\
Three & 50-50 & 19.25 & 14 & 30 & 20 & -- & 0.10 & 7.15 & 0.25\\
Three & 20-80 & 19.75 & 15.50 & 100 & 20 & -- & -- & 1.98 & 0.33\\
Three & 20-80 & 19.50 & 14.75 & 100 & 20 & -- & 0.14 & 4.88 & 0.36\\
Three & 20-80 & 18.25 & 14.75 & 100 & 1 & $-7$ & -- & 1.44 & 0.08\\
Three & 50-50 & 18.00 & 14.75 & 100 & 1 & $-7$ & -- & 1.45 & 0.04\\
All & 50-50 & 19.50 & 15.50 & 40 & 20 & -- & -- & 11.90 & 0.22\\
All & 50-50 & 19.50 & 15.50 & 40 & 20 & -- & 0.08 & 16.06 & 0.22\\
All & 20-80 & 20 & 16 & 100 & 20 & -- & -- & 7.97 & 0.32\\
All & 20-80 & 20 & 18 & 100 & 20 & -- & 0.14 & 10.90 & 0.33\\
\hline
\end{tabular}
\tablefoot{(1) Ratios used in the computation of the best solution with the minimum $\chi^2$. (2) Percentage contribution from illuminated and dark sides of the cloud, respectively. (3) Logarithm of the ionizing flux of the central engine associated with the minimum $\chi^2$ solution. (4) Logarithm of the total pressure of the cloud associated with the minimum $\chi^2$ solution. (5) Microturbulent velocity of the cloud associated with the minimum $\chi^2$ solution. (6) Metallicity of the cloud associated with the minimum $\chi^2$ solution. (7) Extra mechanical heating applied to the cloud, if present. (8) Best fitting covering factor. This parameter is provided only when equivalent width control was used in the computation of the $\chi^2$. (9) Minimum $\chi^2$ associated with the best fitting solution. (10) Covering factor necessary to match the whole {\rm Fe{\sc ii}} emission intensity across the wavelength range of the observed spectrum of RM 102 with the selected {\tt CLOUDY} simulation.}
\label{tab:solutions}
\end{table*}

\subsection{Fits of the three broad band flux ratios with equivalent width control, without mechanical heating}

To ensure sufficient {\rm Fe{\sc ii}} flux while also accurately reflecting the observed ratios, we computed the $\chi^2$ value using a combination of the three ratios and the observed equivalent widths (EWs) of {\rm Fe{\sc ii}} across the four considered spectral ranges under consideration. These calculations were scaled by a constant value of the covering factor $f<1$. In this work, EWs are measured based on the incident continuum. Table \ref{tab:EWtab} lists the observed {\rm Fe{\sc ii}} EWs for the spectral ranges we have considered. 

For our analysis, we used the observed continuum and the observed {\rm Fe{\sc ii}} flux. However, in our simulations, we employed the local incident continuum and the local emission as calculated by {\tt CLOUDY}. In a plane-parallel geometry, the covering factor is set to 1, but we also consider arbitrary value of $f$ to find the best match with the observations. The predicted EWs from {\tt CLOUDY} simulations are compared by automatically selecting values of $f$ that minimize the $\chi^2$.
Figures \ref{fig:contourewz1t0}-\ref{fig:contourewz20t100} in Appendix \ref{ew} show the EWs produced by an array of simulations with varying $Z$ and $V_{\rm turb}$. These figures separately display the EWs produced by the dark side spectrum, the illuminated side spectrum, and the total spectrum, specifically for the UV {\rm Fe{\sc ii}} emission centred on {\rm Mg{\sc ii}}$\lambda2800$. 

When assuming equal contributions from both the dark and bright sides of the cloud, the lowest $\chi^2$ value ($\chi^2 = 7.15$) was found with a parameter set similar to that of the equal contributions case without the EW control, albeit with slightly lower turbulence (see Table \ref{tab:solutions}). In this scenario, the resulting $f=0.10$ minimizes the $\chi^2$. When assuming unequal contributions from both sides, the minimum $\chi^2$ value obtained was 4.88, which, like the previous case, does not differ significantly from the initial solution that did not use EW control.

These results confirm that EW production is broadly in agreement with the trends of the observed ratios, and including EWs in the $\chi^2$ calculation does not drastically change the solutions. However, minimizing $\chi^2$ brings to a substantial underestimation of $f$ with respect to the one obtained by measuring the integrated {\rm Fe{\sc ii}} flux over the entire wavelength range (last column in Table \ref{tab:solutions}) for all cases considered in Table \ref{tab:solutions}.

\begin{table}[h]
\centering
\caption{EW measurements of the interested ranges.}
\begin{tabular}{c c c}
\hline\hline
Spectral range & Wavelength & Observed EW \\
 & [\AA] & [\AA] \\
(1) & (2) & (3) \\
\hline
UV blue wing & 2650 - 2800 & $21.52\pm13.99$ \\
UV red wing & 2800 - 3090 & $55.09\pm22.04$ \\  
Optical blue wing & 4434 - 4684 & $81.52\pm32.61$ \\
Optical red wing & 5150 - 5350 & $79.19\pm30.09$ \\
\hline
\end{tabular}
\tablefoot{(1) Spectral range in which the {\rm Fe{\sc ii}} EW is measured. (2) Wavelength interval corresponding to the spectral range. (3) Measured EW in the selected spectral range, including associated errors.}
\label{tab:EWtab}
\end{table}

\subsection{Solutions with mechanical heating}
\label{solmech}

\begin{figure*}[h]
    \centering
    \includegraphics[width=18.5cm]{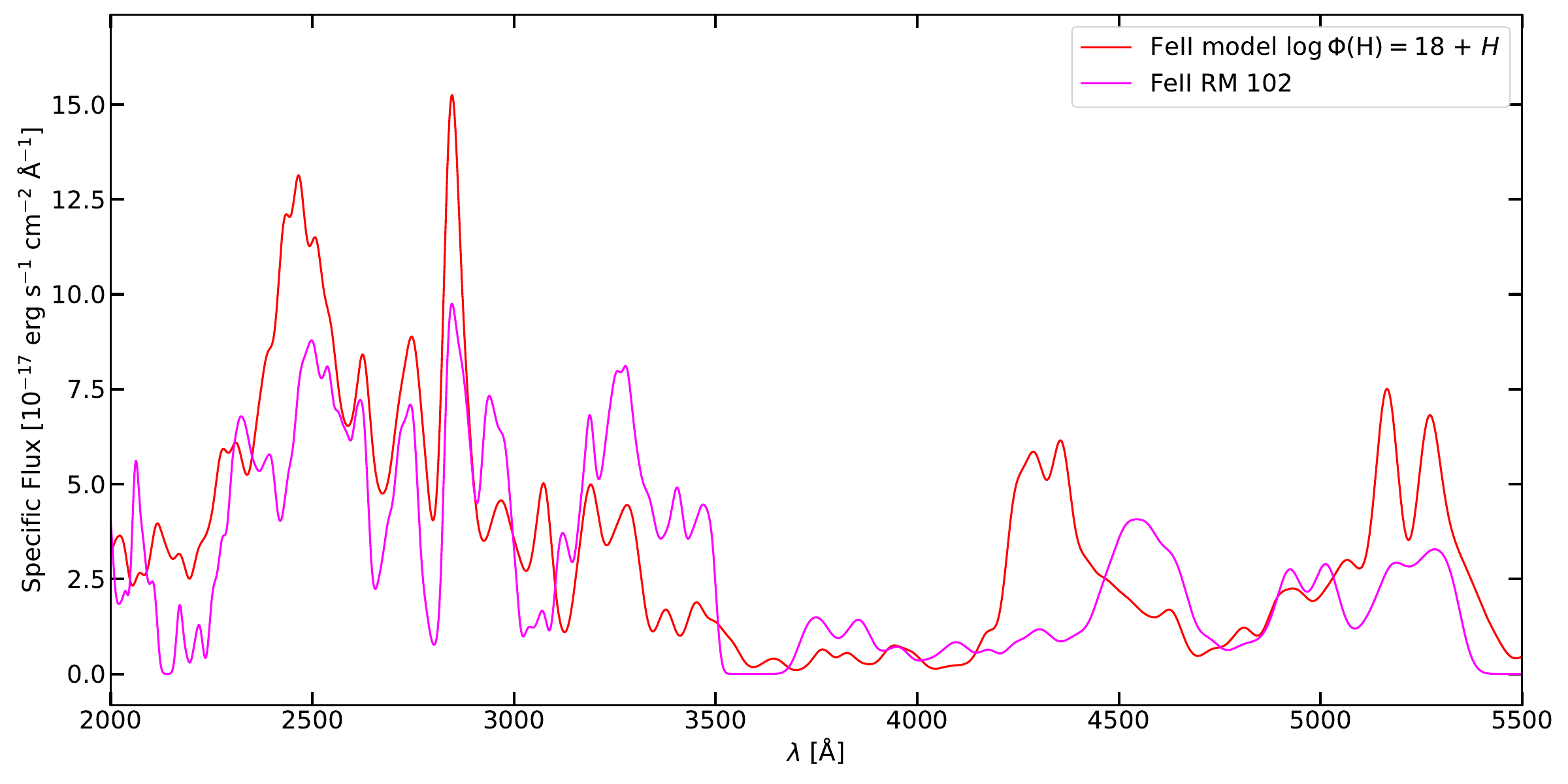}
    \caption{Comparison of the observed {\rm Fe{\sc ii}} emission from the source with the {\rm Fe{\sc ii}} model produced from a {\tt CLOUDY} simulation including mechanical heating (also shown as the red line in Figure \ref{fig:str}), assuming 25\% covering factor.}
    \label{fig:18+h}
\end{figure*}

To investigate the potential contribution of mechanical heating from cloud collisions to the {\rm Fe{\sc ii}} emission, we created a grid of {\tt CLOUDY} simulations with $\log \Phi$(H) ranging from 17 to 22 cm$^{-2}$~s$^{-1}$ and constant pressure covering the $\log (P_{\rm tot})$ range of 14 to 18 cm$^{-3}$ K. The adopted microturbulent velocity for all simulations is $V_{\rm turb} = 100$ km~s$^{-1}$. The additional mechanical heating, as described in Section \ref{mech}, is $H=10^{-7}$ erg cm$^{-3}$ s$^{-1}$. It is important to note, as discussed in Section~\ref{sect:constant_pressure}, that mechanical heating significantly affects the cloud structure and emission only when the ionization flux is low. Figure \ref{fig:18+h} displays a spectrum obtained from a simulation with mechanical heating, demonstrating the enhanced optical {\rm Fe{\sc ii}} emission compared to the observed spectrum of our source.

Using the same approach to calculate the minimum $\chi^2$ from the first three flux ratios for this grid of simulations, and considering both equal and unequal contributions from the illuminated and dark sides of the clouds, we identified the best fits among all {\tt CLOUDY} simulations performed in this study. A shown in Table \ref{tab:solutions}, the two best solutions obtained using this method are very similar, with $\chi^2$ values of 1.44 and 1.45, respectively. The optimal solution is found with parameters of $\log\Phi$(H)$\sim$ 18.25 cm$^{-2}$~s$^{-1}$, $\log (P_{\rm tot})$ = 14.75 cm$^{-3}$ K and $Z = Z_\odot$. 

Despite the remarkably low $\chi^2$ values, indicating a good statistical fit, the resulting ionizing flux in these models is inconsistent with the observed FWHM of the {\rm Fe{\sc ii}} lines. This is because $\Phi$(H) is inversely proportional to the square of the distance to the BLR clouds from the central ionizing source, $r_{\rm BLR}$ (i.e., $\Phi$(H)$\propto r_{\rm BLR}^{-2}$).  Such a low ionizing flux suggests that either extreme shielding is occurring or that the {\rm Fe{\sc ii}} is emitted from a region incompatible with the observed FWHM of {\rm Fe{\sc ii}} and other lines in the spectrum of RM 102. 

Furthermore, as detailed in Section \ref{sect:constant_pressure}, the optical {\rm Fe{\sc ii}} flux is substantially stronger than observed, and EWs would require an unphysically low covering factor of $f<0.10$. The combination of high microturbulence and mechanical heating thus results in an overproduction of {\rm Fe{\sc ii}} that does not align with physical expectations.
Therefore, even though the minimum $\chi^2$ values are low, suggesting a good fit statistically, no solution incorporating mechanical heating is considered physically plausible for this scenario.

\subsection{Contribution from individual transitions}
\label{transitions}

In Section~\ref{sect:three_ratios}, we demonstrated that the solutions obtained by considering the first three flux ratios do not fully reproduce the observed spectrum when using the adopted covering factors. Several spectral features exhibited significant discrepancies between the observed data and the model, affecting the shape of the emission template. In this section, we delve deeper into these discrepancies by examining the intensities of specific transitions as functions of the simulation parameters.

Following the work of \cite{kocacevic2015} and \citet{popovic2019}, we focus on UV multiplets 60, 61, 62, 63, 78, along with two empirical transitions observed in the I~Zw~1 UV spectrum from \cite{popovic2019}. These transitions were centred at 2720 \AA~ and 2840 \AA, respectively, which we refer to as IZw1a and IZw1b. \citet{popovic2019} suggested that these transitions likely originate from high excitation levels not included in their study. However, these transitions are correctly incorporated in the {\tt CLOUDY} simulations used here and are present in the spectra generated with our preferred set of parameters outlined in Section \ref{sect:three_ratios}. Additionally, we analyze the spike-to-gap ratio, a key diagnostic of plasma conditions \citep{baldwin2004,sarkar2021}. The specific spectral ranges associated with each of these transitions are described in Table~\ref{table:UV_lines}.

To improve the template fits, we incorporated intensity ratios of these specific transitions into the $\chi^2$ computation, using them as diagnostics to refine the accuracy of our models. These ratios, which we refer to as the Spike-to-spike ratio, the Spike-to-gap ratio \citep{sarkar2021}, $R_{2800}$ and $R_{2900}$, are listed in Table~\ref{tab:measuredratios} along with their observed values in the spectrum of RM 102. 

\begin{figure*}[h]
    \centering
    \includegraphics[width=18.5cm]{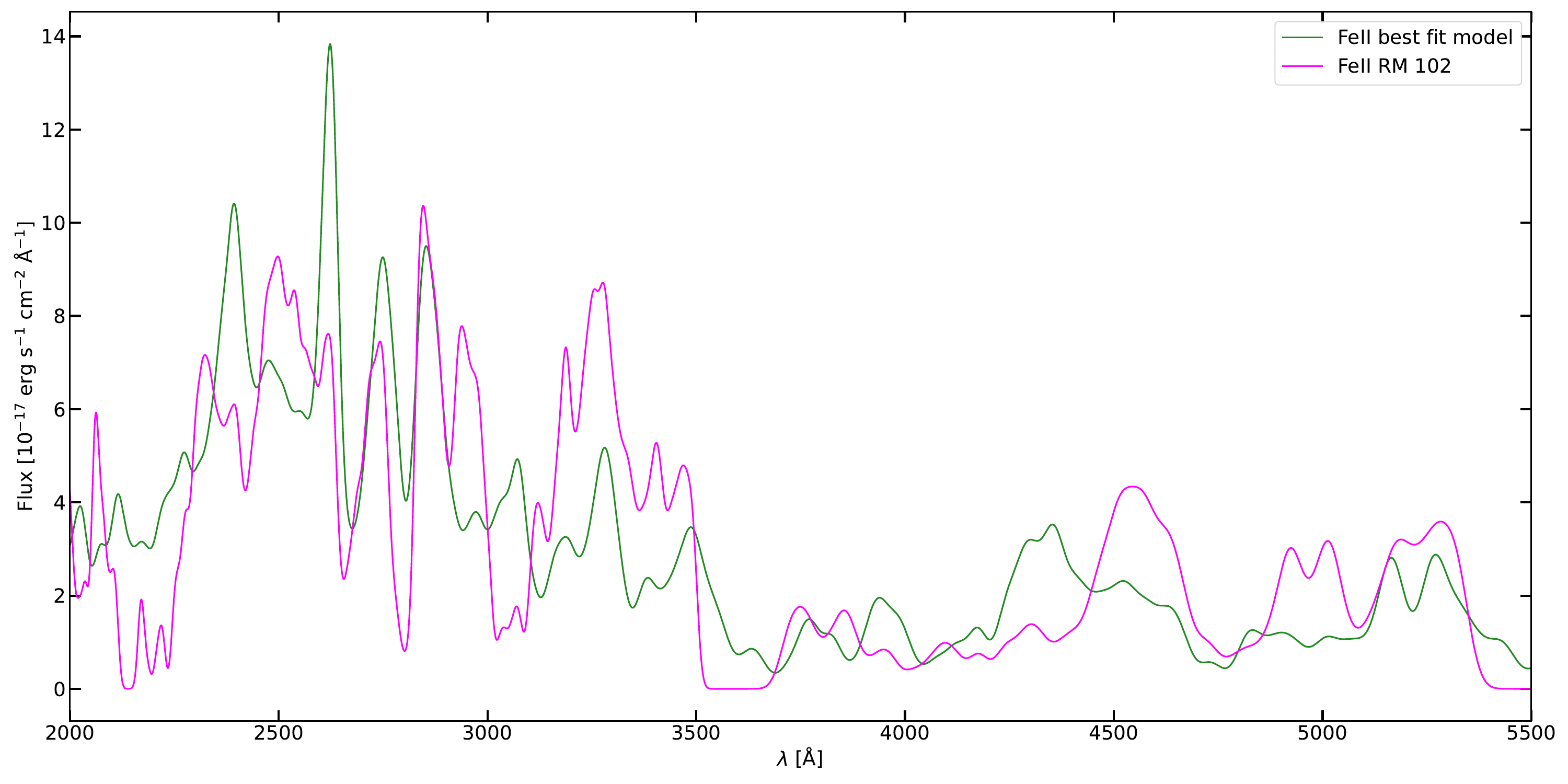}
    \caption{{\rm Fe{\sc ii}} spectrum assuming a 20\% contribution from bright sides and 80\% contribution from dark sides with minimum $\chi^2$ employing the ratios in Section \ref{sect:fitting}, displayed with 33\% covering factor. The {\rm Fe{\sc ii}} template corresponds to the {\tt CLOUDY} simulation associated with $\log\Phi$(H) = 20 cm$^{-2}$~s$^{-1}$, $\log(P_{\rm tot}) = 16$ cm$^{-3}$ K, $Z=20 Z_\odot$ and $V_{turb} = 100$ km~s$^{-1}$.}
    \label{fig:bestfit}
\end{figure*}

After including these additional ratios to the $\chi^2$ calculation, the best-fitting model, with $\chi^2=7.97$, corresponds to a reasonable set of parameters: $\log\Phi$(H)=20 cm$^{-2}$~s$^{-1}$, $\log(P_{\rm tot})=16$ cm$^{-3}$ K, $V_{\rm turb}=100$ km s$^{-1}$ and $Z=20Z_\odot$. This model assumes unequal contributions from the bright and dark sides of the cloud. The spectrum produced by this simulation is shown in Figure \ref{fig:bestfit}. Although the $\chi^2$ value increases slightly when including the four additional ratios, the match with the observed data does not deteriorate. On the contrary, the template generated from this specific simulation, using the covering factor detailed in Table \ref{tab:solutions}, aligns closely with most of the features observed in the original spectrum of RM 102. The primary exceptions are of the two strong spikes associated with the spike-to-spike and the spike-to-gap ratios, which strongly increase in intensity with higher metallicities (see Appendix \ref{app:other_ratios}), and the features at 4350\AA\ as well as the weaker optical blue wing of {\rm Fe{\sc ii}}. Despite these discrepancies, this simulation represents our preferred solution because it achieves a good balance between fitting the observed data and maintaining a physically plausible set of parameters, requiring minimal shielding and suggesting a cloud density of $n_{\rm H}\sim 10^9$ cm$^{-3}$.

The contour plots for the additional ratios are shown in Appendix~\ref{app:other_ratios}. Interestingly, our constant pressure analysis of the spike-to-gap ratio (see Appendix~\ref{app:other_ratios}) favours the turbulent velocity of 40 km~s$^{-1}$ (consistent with the requirement for three basic ratios) but indicates a lower metallicity, solar to 5 times solar.


\begin{table}[h]
\caption{UV {\rm Fe{\sc ii}} lines}
\label{table:UV_lines}
\centering
\begin{tabular}{c c}
\hline\hline
{\rm Fe{\sc ii}} UV line & Spectral range\\ 
(1) & (2)\\
\hline
Spike$_1$ & $2280$-$2430$ \AA\\
Spike$_2$ & $2560$-$2660$ \AA\\
Gap & $2430$-$2560$ \AA\\
UV60 & $2907$-$2980$ \AA\\
UV61 & $2833$-$2918$ \AA\\
UV62 & $2692$-$2756$ \AA\\
UV63 & $2614$-$2773$ \AA\\
UV78 & $2944$-$3003$ \AA\\
IZw1a & $2695$-$2735$ \AA\\
IZw1b & $2820$-$2860$ \AA\\
\hline
\end{tabular}
\tablefoot{(1) UV {\rm Fe{\sc ii}} line considered. (2) Spectral range considered for the estimation of the intensity of the line.}
\end{table}

\begin{table}[h]
\renewcommand{\arraystretch}{1.5}
\caption{UV {\rm Fe{\sc ii}} line ratios measured in the spectrum of RM 102}
\label{tab:measuredratios}
\centering
\begin{tabular}{l c r}
\hline\hline
Name & Ratio & Value\\ 
(1) & (2) & (3)\\
\hline
Spike-to-spike & $\frac{{\rm Spike}_2}{{\rm Spike}_1}$ & $0.709\pm0.369$\\
Spike-to-gap & $\frac{{\rm Spike}_1 + {\rm Spike}_2}{\rm Gap}$ & $1.438\pm0.647$\\
$R_{2800}$ & $\frac{{\rm IZw1b}}{\rm UV63}$ & $0.913\pm0.703$\\
$R_{2900}$ & $\frac{{\rm IZw1b} + {\rm UV61}}{{\rm UV60} + {\rm UV78}}$ & $1.175\pm0.494$\\
\hline
\end{tabular}
\tablefoot{(1) Name of the ratio. (2) Expression for the measurement of the ratio. (3) Observed value of the ratio in the spectrum of RM 102, with error.}
\end{table}



\section{Discussion}
\label{discuss}

In this study, we modelled the {\rm Fe{\sc ii}}  contribution to the emission of RM 102, a quasar that is highly representative of the general quasar population as well as xA sources as shown in Figure \ref{fig:fig1}. Our model comprehensively covers both the UV and optical bands, utilizing the latest atomic data available in the current version of CLOUDY \citet{Cloudy23} and incorporating all available transitions. The modeling was based on the assumption of constant pressure, which is more appropriate for maintaining cloud stability. We explored a broad range of parameters, including ionization flux, total pressure, metallicity, and turbulent velocity. Additionally, we considered the effects of mechanical heating and the geometrical enhancement of the contribution from the dark sides of the clouds to the total spectrum. The most relevant fits are summarized in Table~\ref{tab:solutions}. There, the {\rm Fe{\sc ii}} emission is measured with respect to the incident continuum assumed in the model.

Although our simulations do not fully reproduce the observed spectrum, they allow us to draw several tentative conclusions, which are discussed below.

\subsection{Constant density vs. constant pressure}

In Section \ref{turbdp}, we discussed the role of turbulent velocity in both constant pressure and constant density simulations. Turbulent velocity, along with metallicity, significantly influences the parameter space of $\Phi$(H) and $P_{\rm tot}$, as observed in the contour plots in Figures \ref{fig:turbplots5050} and \ref{fig:turbplots2080}. This is the primary distinction from the approach taken by \cite{Pandey2024}, which assumes constant density.  However, the constant pressure assumption more effectively captures the variations in the physical conditions of the gas driven by changes in metallicity and turbulent velocity. This approach allows the observed ratios to explore a broader range of the parameter space. Incorporating turbulent velocity ultimately leads to larger cloud sizes and more complex cloud structures.

\subsection{Consistency of the required and observed $\Phi$(H), and the potential need for shielding}
\label{undcont}

The apparent overproduction of the {\rm Fe{\sc ii}} UV emission in our models might be due to inaccuracies in determining the underlying continuum. This might be caused still by some confusion with pseudocontinua, with reddening, and with the starlight contribution. his issue could stem from confusion with pseudocontinua, reddening effects, or contributions from starlight. The mean slope of the spectrum above $\sim 5000$ \AA~is 2.08, which is only slightly flatter than the slope of 2.33 expected from a standard accretion disk \citep{SS1973}. This flattening is likely due to starlight contribution, as RM 102 is not particularly luminous, with a black hole mass just above $10^8 M_{\odot}$. The modest luminosity leaves little room for significant internal reddening. 

We also explored the possibility of reddening by examining the Balmer decrement. Although the higher-order Balmer lines are clearly visible in the spectrum, the line ratios are quite uncertain (see Table~\ref{table:Balmer}). This uncertainty is likely because Balmer line ratios are sensitive to both the density and temperature of the clouds, as indicated by both simulations and observational data \citep{ilic2012}. Nevertheless, the observed ratios are consistent with theoretical expectations \citep{Osterbrock1989} and with the typical values for AGN as reported in \citet{popovic2003}. This consistency supports the conclusion that the disk emission in RM 102 is not significantly affected by intrinsic reddening.

Performed computations did not rely on the precise distance measurement from the SMBH to the BLR. However, this distance, inferred from the time delay between the continuum and line emissions, has been estimated and is reported in Table~\ref{tab:RM102_info}. This delay implies a specific value for the ionizing flux. We observe that the solutions presented in Table~\ref{tab:solutions} do not fully satisfy this condition, with some deviating significantly from the values based on the observed continuum and the measured time delay. This discrepancy could imply that the net time delay does not accurately reflect the distribution of the flux within the source, or that the BLR actually experiences a different incident continuum than the one observed, potentially due to some form of obscuration. Both effects could be significant, but the substantial mismatch between the measured and modeled values of $\Phi$(H) is concerning, particularly since the FWHMs of the {\rm Fe{\sc ii}} and H$\beta$ lines are very similar, as discussed in Section~\ref{solmech}. It is challenging to disentangle these effects reliably. 

Without a precise distance, we cannot scale the model to the data in absolute units. If we assume the delay provides the correct distance, then the strength of the features (effectively, the covering factor) reported in Table~\ref{tab:solutions} should be rescaled to match the observed continuum by accounting for the difference in assumed and observed $\Phi$(H). This adjustment would immediately favor the last two models, which are closest to the observationally inferred $\Phi$(H).

With the exception of the best solutions obtained with the addition of mechanical heating in the simulations, most of the best solutions found in this work (see Table \ref{tab:solutions}) require only marginal shielding relative to the measured value of $\log\Phi$(H)$ \sim 20.5$ cm$^{-2}$~s$^{-1}$ of our source. This consistency is noteworthy, as it suggests that substantial shielding is not necessary for our favored model, which minimizes $\chi^2$ by considering all ratios and assumes a 20\% contribution from the bright side and an 80\% contribution from the dark side of the cloud.


\begin{table}[h]
\caption{Balmer decrement in RM 102}
\label{table:Balmer}
\centering
\begin{tabular}{c c c c}
\hline\hline
Ratio & Observed & Predicted & AGN sample\\ 
(1) & (2) & (3) & (4) \\
\hline
H$\gamma$/H$\beta$ & $0.60\pm0.36$ & 0.47 & 0.45\\
H$\delta$/H$\beta$ & $0.20\pm0.09$ & 0.26  & 0.21\\
\hline
\end{tabular}
\tablefoot{(1) Balmer decrement ratio. (2) Observed value of the ratio measured on the spectrum of RM 102. (3) Predicted value of the ratio in the absence of extinction \citep{Osterbrock1989}. (4) Average values measured for an AGN sample by \citet{popovic2003} for comparison.}
\end{table}



\subsection{Are dark sides of clouds favoured?}

Examining the contours in Figure \ref{fig:turbplots2080}suggests a strong preference for an inflow pattern geometry, where the dark sides of the clouds are disproportionately represented compared to their illuminated sides. Assuming $V_{\rm turb} \sim 100$ km~s$^{-1}$ (non-thermal turbulence, consistent with the chaotic motion of clouds), we find that although the contours do not perfectly overlap, this configuration provides the best fit. It maintains reasonable physical conditions that align with both our models and observations. The optical ratio still tends towards higher values of $\Phi$(H); however, within error, these ratios can be consistent with canonical parameters associated with the physical conditions of the BLR \citep{netzer2013, netzer2020}, or potentially with slightly higher values of $V_{\rm turb}$. 

It is noteworthy that, although the microturbulence parameter cannot be directly measured in the source, $V_{\rm turb}\sim 100$ km~s$^{-1}$ is not extreme and has been adopted in previous studies \citep{panda_etal_2018,panda_etal_2019,sarkar2021, panda2021, Pandey2024}. Slightly higher values of $V_{\rm turb}$ could potentially reconcile all three ratios together, though a combination with high $Z$ could potentially lead to significant overproduction of {\rm Fe{\sc ii}} emission.

In conclusion, when comparing the contours in Figures \ref{fig:turbplots5050} and \ref{fig:turbplots2080}, the difference is striking. Using only the first three diagnostic ratios studied in Section \ref{sect:three_ratios}, it is evident that the only way to reproduce physical conditions consistent with the observed $\Phi$(H) and appropriate density values for the BLR, is assuming  $Z\sim20Z_\odot$, $V_{\rm turb}\sim100$ km~s$^{-1}$ and a geometry that favours dark sides. This trend is also observed when considering all the additional ratios discussed in Section \ref{transitions}.

\subsection{The role of mechanical heating in the BLR}

Mechanical heating does improve the best $\chi^2$ with just a handful of simulations needed. The best $\chi^2$ values found are associated with simulations featuring $\Phi$(H) levels that are incompatible with the observations of RM 102. In both scenarios --whether assuming equal or unequal contributions from the bright and dark sides of clouds-- the optimal $\Phi$(H) is incompatible with the observed FWHM of {\rm Fe{\sc ii}} lines, as well as other emission lines. These simulations would require extreme levels of internal reddening or shielding that are not observed. Therefore, the inclusion of mechanical heating does not yield a significant improvement in modeling {\rm Fe{\sc ii}} emission, nor does it enhance the compatibility across different diagnostic ratios.

\subsection{Model constraints for the density, turbulence and metallicity}

All the best-fit solutions presented in Table \ref{tab:solutions} indicate a general trend of average $n_{\rm H}\sim 10^9$ cm$^{-3}$, with the exception of the solution with $\log(P_{\rm tot})=18$ cm$^{-3}$ K, which exhibits a higher density of $n_{\rm H}\sim 10^{11}$ cm$^{-3}$. Solutions derived using the minimum $\chi^2$ method strongly favour a condition of $n_{\rm H} \approx 10^9$ cm$^{-3}$ in the BLR. These lower density values for RM 102 could be further validated by measuring the intensity of the {\rm C{\sc{iii]}}} emission, which is not present in the spectrum used in this study. 
There is also a consistent trend in the preferred solutions regarding microturbulence, with $V_{\rm turb}\sim 100$ km s$^{-1}$. The metallicity is expected to lie within the range of $5Z_\odot \leq Z \leq 20Z_\odot$. Solutions with higher metallicity are not consistent with our findings for this source and are thus excluded. 



\subsection{Multiple emission regions}

A single cloud approximation does not perfectly match the overall spectrum of RM 102. 
It was frequently suggested that optical and UV {\rm Fe{\sc ii}} emission originate from two distinct regions \citep{baldwin2004, gaskell2022}. Some correlation between the two regions is observed \citep[e.g][]{kocacevic2015, panda_etal_2019}, and the measured time delays of the optical and UV emission are not significantly different as shown by \citet{zajacek24}. However, the significant average redshift in the UV {\rm Fe{\sc ii}} lines, which is not seen in the optical UV counterparts, suggests an outflow component predominantly affecting the UV region \citep{kocacevic2015}. Additionally, \citet{gaskell2022} suggests that the optical {\rm Fe{\sc ii}} is subject to considerable reddening, which is not observed in the UV {\rm Fe{\sc ii}} emission. Given these discrepancies, we tested whether allowing for two distinct types of clouds could improve the representation of the data.

To explore this, we combined emissions from two clouds and calculated the minimum $\chi^2$ among all combinations of simulations. Despite this effort, all automatic fitting processes converged back to a single-cloud model, reproducing the results obtained for scenarios where both the bright and dark sides of the clouds contributed equally and for scenarios emphasizing dark side emissions. Thus, no combination of simulations appears to enhance the single-cloud interpretation.

This approach, however, does not fully address the common mismatch between optical and UV observational templates that are applied to the same sources with different normalizations. In reality, no simulation achieves a perfect optical shape and intensity while simultaneously producing no UV emission, or vice versa for the optical range. Consequently, models that consider only optical or only UV emissions as originating from two physically distinct regions cannot provide a complete solution because the radiation from each region inevitably affects the other band. Typically, our simulations generate similar amounts of UV emission but show wide variation in the optical range, especially when additional heating is included. Conversely, the ability of {\rm Fe{\sc ii}} observational templates based on IZw1 to fit the spectra of most quasars remains unexplained, while theoretical templates fail to match the optical wavelength range when typical BLR physical conditions are assumed \citep{gaskell2022, zhang2024}.

This is likely to be the case for any multi-zone modeling approach, such as the Locally Optimized Cloud (LOC) model proposed by \citet{baldwin1995} or any Monte Carlo-type approach combining multiple emitting regions. Specifically, spectral features that are not well-modeled by any single solution, despite representing a broad parameter space, are unlikely to be accurately reproduced by combining multiple solutions. However, we did not pursue such an approach in the current work.




\subsection{Disk winds possibly shielding the bright sides}

The predominance of emission from the dark sides of the clouds suggests that there may be some form of shielding preventing the bright sides, which are located further from the observer and behind the black hole, from significantly contributing to the observed spectrum. A similar effect has been observed in reverberation mapping studies of certain sources, where the response function peaks at the shortest possible time delay, close to the observer’s line of sight. In these cases, a second peak associated with the bright sides is present but shows a much-reduced amplitude \citep[see e.g.][for NGC 5548]{horne2021}. However, we do not observe substantial reddening of the disk in RM 102, as the spectral slope is roughly consistent with that of a standard accretion disk (see Sect.~\ref{sect:fitting}). Attempts to introduce intrinsic reddening using the extinction law from  \cite{czerny2004} did not improve the spectral fit, indicating that any medium shielding the bright sides is likely dust-free and highly ionized.

Such a medium could plausibly be in the form of magnetically-driven winds that originate close to the black hole \citep[e.g.][and the references therein]{fukumura2024}. These outflows are challenging to detect directly, but there is observational evidence of ultra-fast outflows (UFOs) with velocities ranging from 0.03 to 0.6 times the speed of light. These UFOs are detected through shifts in the emission lines of highly ionized species in the X-ray band \citep[e.g.][]{gianolli2024}. The existence of such winds could provide a natural explanation for the observed shielding effects.

\subsection{A comment on the current state of atomic transitions}

Some observed spectral features, such as the dip near 2400\AA~ and the additional feature around $4350\AA$, are not reproduced in our models. {\tt CLOUDY} offers various atomic databases for use in simulations, and the differences between these databases have been discussed by \citet{sarkar2021}. For this study, we utilized the {\rm Fe{\sc ii}} database from \citet{Smyth2019}, as recommended by \citet{sarkar2021}. This database is now the default in {\tt CLOUDY} and includes a comprehensive set of atomic data, representing the most advanced modeling currently available. However, the discrepancies between predicted and observed features suggest that there is still significant room for improvement in the atomic data and modeling techniques.

Additionally, the methods used to solve the radiative transfer equations may contribute to these discrepancies. \citet{netzer2020} pointed out that current photoionization codes using the escape probability formalism often fail to accurately compute transitions in relatively dense, optically thick clouds. While alternative codes such as {\tt TITAN} \citep{Dumont2000, dumont2003} and {\tt TLUSTY} \citep{hubeny2000} exist, they lag behind {\tt CLOUDY} in terms of the richness of their atomic data, even in their most recent versions \citep[e.g.][]{palit2024}. This highlights the need for further development in both atomic databases and radiative transfer methodologies to enhance the accuracy of modeling complex astrophysical environments.

\section{Conclusions}
\label{conclusions}

After analyzing the {\rm Fe{\sc ii}} emission of the bright quasar RM 102 employing a range of techniques and assumptions, we draw several key conclusions.

\begin{itemize}

\item We find strong evidence supporting the predominance of the dark sides of clouds over isotropic emission. This finding aligns with previous studies on the H$\beta$ profile using reverberation mapping \citep{horne2021}. The similarity in the FWHM of the {\rm Fe{\sc ii}} lines and H$\beta$ confirms that {\rm Fe{\sc ii}} originates from the same regions as H$\beta$, underscoring the critical role of dark sides in {\rm Fe{\sc ii}} emission. This predominance is crucial for matching the observed UV/optical ratios and the UV red-to-blue ratios in the $\Phi$(H)-$n_{\rm H}$ parameter space compatible with the expected physical conditions in the BLR \citep{gaskell2022} and the observations of RM 102.

\item Our results indicate significant chemical enrichment in the BLR of RM 102. Combined with high turbulent velocities and the predominance of dark sides, this enrichment creates the necessary conditions to produce most of the observed {\rm Fe{\sc ii}} emission ratios. The high metallicity inferred from our models is consistent with findings from similar studies of other quasars \citep{nagao2006, maiolino2019,panda_etal_2019_QMS, sniegowska2021, panda2021, garnica2022, floris2024}.

\item The favoured solution described in Section \ref{transitions} with the minimum $\chi^2$ result, of $\log\Phi$(H)=20 cm$^{-2}$ s$^{-1}$, $Z=20Z_\odot$, $\log (P_{\rm tot})$= 16 cm$^{-3}$ K and $V_{\rm turb}= 100$ km s$^{-1}$, aligns well with the measured $\Phi$(H) from RM 102. This solution also supports the presence of low-density ($n_{\rm H}\sim 10^9$ cm$^{-3}$) BLR clouds. Although this density is lower than typical values for BLR clouds, observations of {\rm C{\sc{iii]}}} intensity with instruments such as X-Shooter at the Very Large Telescope (VLT) could provide further insight into the low density observed and help refine these constraints.

\item While some discrepancies in observed {\rm Fe{\sc ii}} features might suggest multiple emission regions, our study found no compelling evidence to support this hypothesis. Further analysis using more advanced multi-zone modelling approaches, such as the Locally Optimized Cloud (LOC) method \citep{baldwin1995} or Monte Carlo-type simulations, could offer additional insights into this possibility.

\item The absence of specific {\rm Fe{\sc ii}} features in the observed spectra of RM 102, as well as the appearance of additional features in the simulations (such as the emission at 4350 \AA), along with notable discrepancies in the intensity of specific lines, highlight potential limitations in the atomic data used in {\tt CLOUDY}. Despite being one of the most comprehensive spectral synthesis codes, there may be missing transitions or inaccuracies in transition rates affecting the modelling of {\rm Fe{\sc ii}} emission. Further research on {\rm Fe{\sc ii}} emission is necessary to address these gaps.

\end{itemize}



\begin{acknowledgements}
This project has received funding from the European Research Council (ERC) under the European Union’s Horizon 2020 research and innovation program (grant agreement No. [951549]). BC acknowledges the OPUS-LAP/GA CR-LA bilateral project (2021/43/I/ST9/01352/OPUS
22 and GF23-04053L). AF acknowledges financial support from the Center for Theoretical Physics, Polish Academy of Sciences during his stay in Warsaw. AP acknowledges funding from the Chinese Academy of Sciences President’s International Fellowship Initiative (PIFI), Grant No. 2024PVC0088. MLM-A acknowledges financial support from Millenium Nucleus NCN19-058 and NCN2023${\_}$002 (TITANs) an ANID Millennium Science Initiative (AIM23-0001). SP acknowledges the financial support of the Conselho Nacional de Desenvolvimento Científico e Tecnológico (CNPq) Fellowship 301628/2024-6 and is supported by the international Gemini Observatory, a program of NSF NOIRLab, which is managed by the Association of Universities for Research in Astronomy (AURA) under a cooperative agreement with the U.S. National Science Foundation, on behalf of the Gemini partnership of Argentina, Brazil, Canada, Chile, the Republic of Korea, and the United States of America.
\end{acknowledgements}

\bibliographystyle{aa}
\bibliography{biblio.bib}

\begin{appendix} 

\section{Minimum $\chi^2$ solutions}
\label{app:minchifig}

\begin{figure*}[h]
    \centering
    \includegraphics[width=18.5cm]{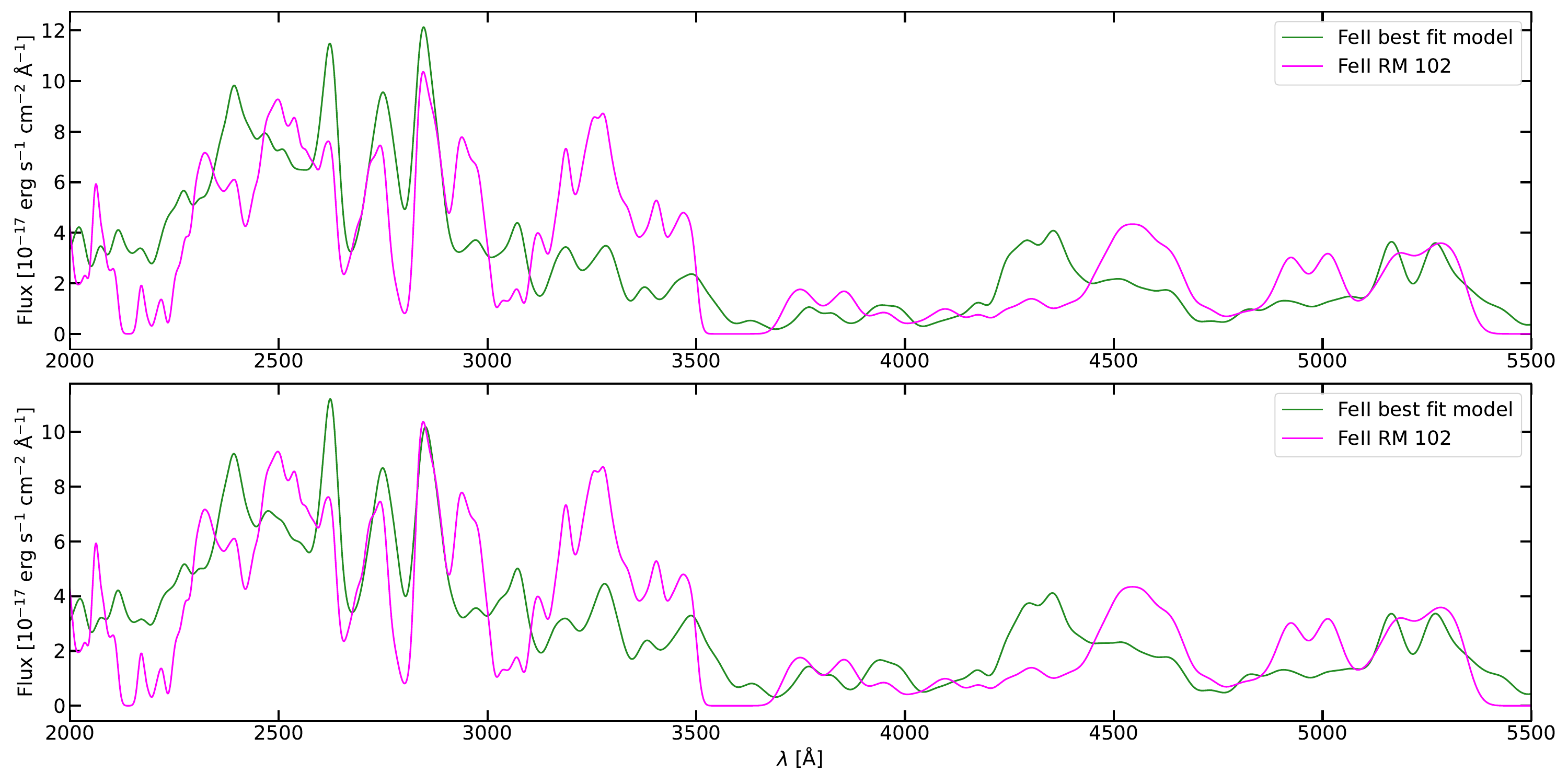}
    \caption{Comparison of the {\rm Fe{\sc ii}} spectrum from our source with the best $\chi^2$ solutions employing the first three ratios (Section \ref{sect:fitting}) in the computation. {\it Top}: {\rm Fe{\sc ii}} spectrum assuming equal contribution from bright and dark sides, displayed with a 24\% covering factor. The {\rm Fe{\sc ii}} template corresponds to the {\tt CLOUDY} simulation with $\log\Phi$(H) = 19.25 cm$^{-2}$~s$^{-1}$, $\log(P_{\rm tot}) = 14$ cm$^{-3}$ K, $Z=20 Z_\odot$ and $V_{turb} = 40$ km~s$^{-1}$. {\it Bottom}: {\rm Fe{\sc ii}} spectrum assuming a 20\% contribution from bright sides and 80\% contribution from dark sides, displayed with a 33\% covering factor. The {\rm Fe{\sc ii}} template corresponds to the {\tt CLOUDY} simulation associated with $\log\Phi$(H) = 19.75 cm$^{-2}$~s$^{-1}$, $\log(P_{\rm tot}) = 15.50$ cm$^{-3}$ K, $Z=20 Z_\odot$ and $V_{turb} = 100$ km~s$^{-1}$.}
    \label{fig:best3}
\end{figure*}

\begin{figure*}[h]
    \centering
    \includegraphics[width=18.5cm]{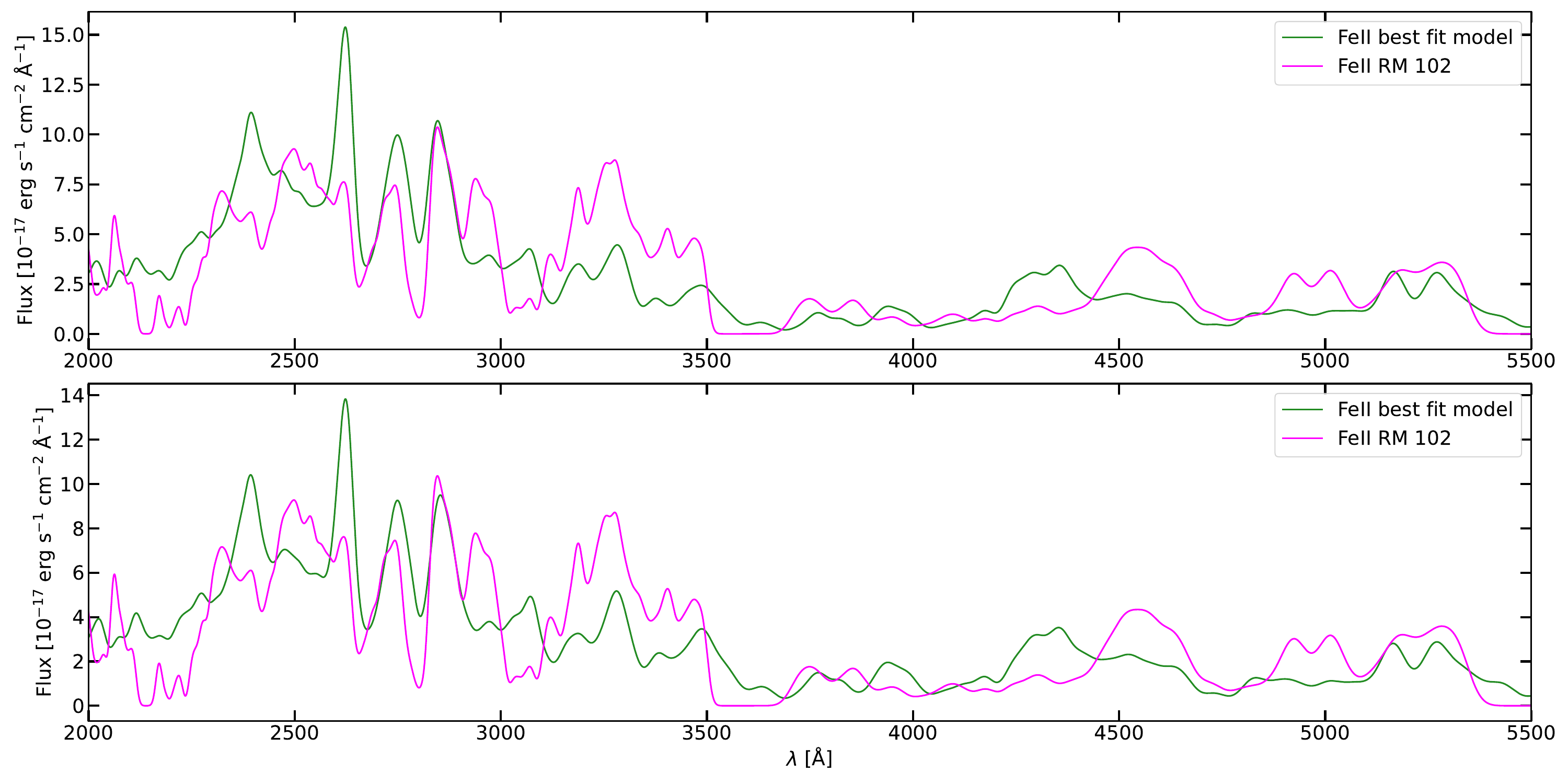}
    \caption{Same as in Figure \ref{fig:best3}, but for the best $\chi^2$ solutions obtained by also employing the ratios discussed in Section \ref{transitions}. {\it Top}: {\rm Fe{\sc ii}} spectrum assuming equal contribution from bright and dark sides, displayed with a 22\% covering factor. The {\rm Fe{\sc ii}} template corresponds to the {\tt CLOUDY} simulation associated with $\log\Phi$(H) = 19.50 cm$^{-2}$~s$^{-1}$, $\log(P_{\rm tot}) = 15.50$ cm$^{-3}$ K, $Z=20 Z_\odot$ and $V_{turb} = 40$ km~s$^{-1}$. {\it Bottom}: {\rm Fe{\sc ii}} spectrum assuming a 20\% contribution from bright sides and 80\% from dark sides, displayed with a 32\% covering factor. The {\rm Fe{\sc ii}} template corresponds to the {\tt CLOUDY} simulation associated with $\log\Phi$(H) = 20 cm$^{-2}$~s$^{-1}$, $\log(P_{\rm tot}) = 16$ cm$^{-3}$ K, $Z=20 Z_\odot$ and $V_{turb} = 100$ km~s$^{-1}$.}
    \label{fig:bestall}
\end{figure*}

\begin{figure*}[h]
    \centering
    \includegraphics[width=18.5cm]{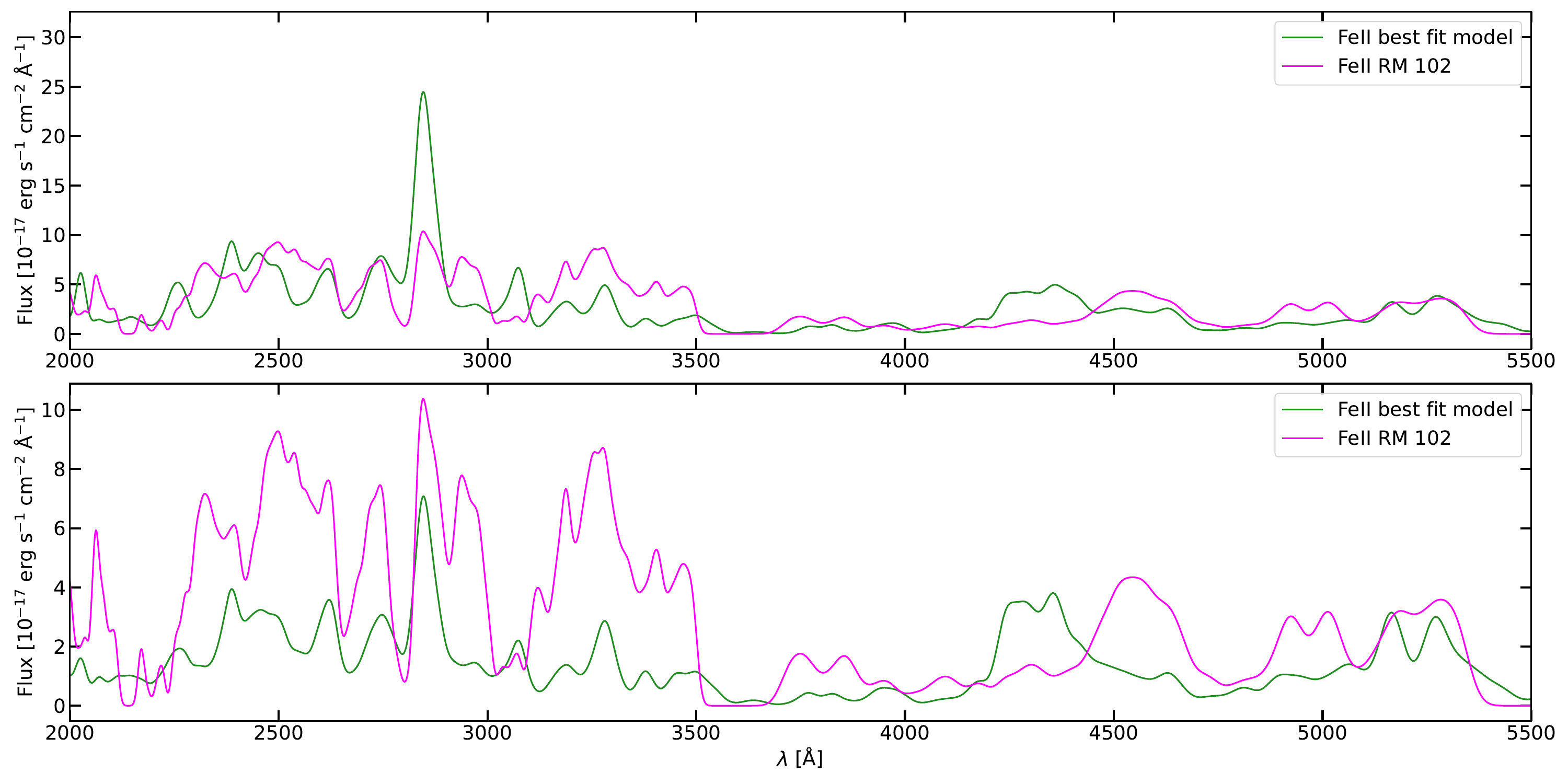}
    \caption{Same as in Figure \ref{fig:best3}, but for the best $\chi^2$ solutions obtained using simulations that include mechanical heating as discussed in Section \ref{solmech}. {\it Top}: {\rm Fe{\sc ii}} spectrum assuming equal contribution from bright and dark sides, displayed with a 4\% covering factor. The {\rm Fe{\sc ii}} template corresponds to the {\tt CLOUDY} simulation associated with $\log\Phi$(H) = 18 cm$^{-2}$~s$^{-1}$, $\log(P_{\rm tot}) = 14.75$ cm$^{-3}$ K, $Z=Z_\odot$ and $V_{turb} = 100$ km~s$^{-1}$. {\it Bottom}: {\rm Fe{\sc ii}} spectrum assuming a 20\% contribution from bright sides and 80\% from dark sides, displayed with an 8\% covering factor. The {\rm Fe{\sc ii}} template corresponds to the {\tt CLOUDY} simulation associated with $\log\Phi$(H) = 18.25 cm$^{-2}$~s$^{-1}$, $\log(P_{\rm tot}) = 14.75$ cm$^{-3}$ K, $Z=Z_\odot$ and $V_{turb} = 100$ km~s$^{-1}$.}
    \label{fig:bestmech}
\end{figure*}

Figures \ref{fig:best3} to \ref{fig:bestmech} display the spectra corresponding to the template solutions with the minimum $\chi^2$ values, as detailed in Table \ref{tab:solutions}. These figures present the best-fitting models for each scenario explored in our analysis, excluding those solutions with EW control, which generally match the results obtained without employing the EW in the $\chi^2$ computation.

\section{Ratio trends}
\label{app:ratios}

Figures \ref{fig:threeratio_trends0} to \ref{fig:Ratio_trends100_2080} display the trends associated with the ratios considered in this work. Each figure shows trends for different combinations of $V_{\rm turb}$, $Z$, and varying contributions from bright and dark sides of the clouds. The red contour (when present) represents the exact value of the considered ratio, as measured in the object RM 102.

\subsection{Trends of canonical ratios}
\label{app:3ratios}

We first present the trends associated with the canonical ratios described in Section \ref{sect:fitting} within the selected parameter space of $\log\Phi$(H), $\log(P_{\rm tot})$, $V_{\rm turb}$ and $Z$.

\subsubsection{UV-to-optical ratio}

\begin{figure*}[h]
    \centering
    \includegraphics[width=18.5cm]{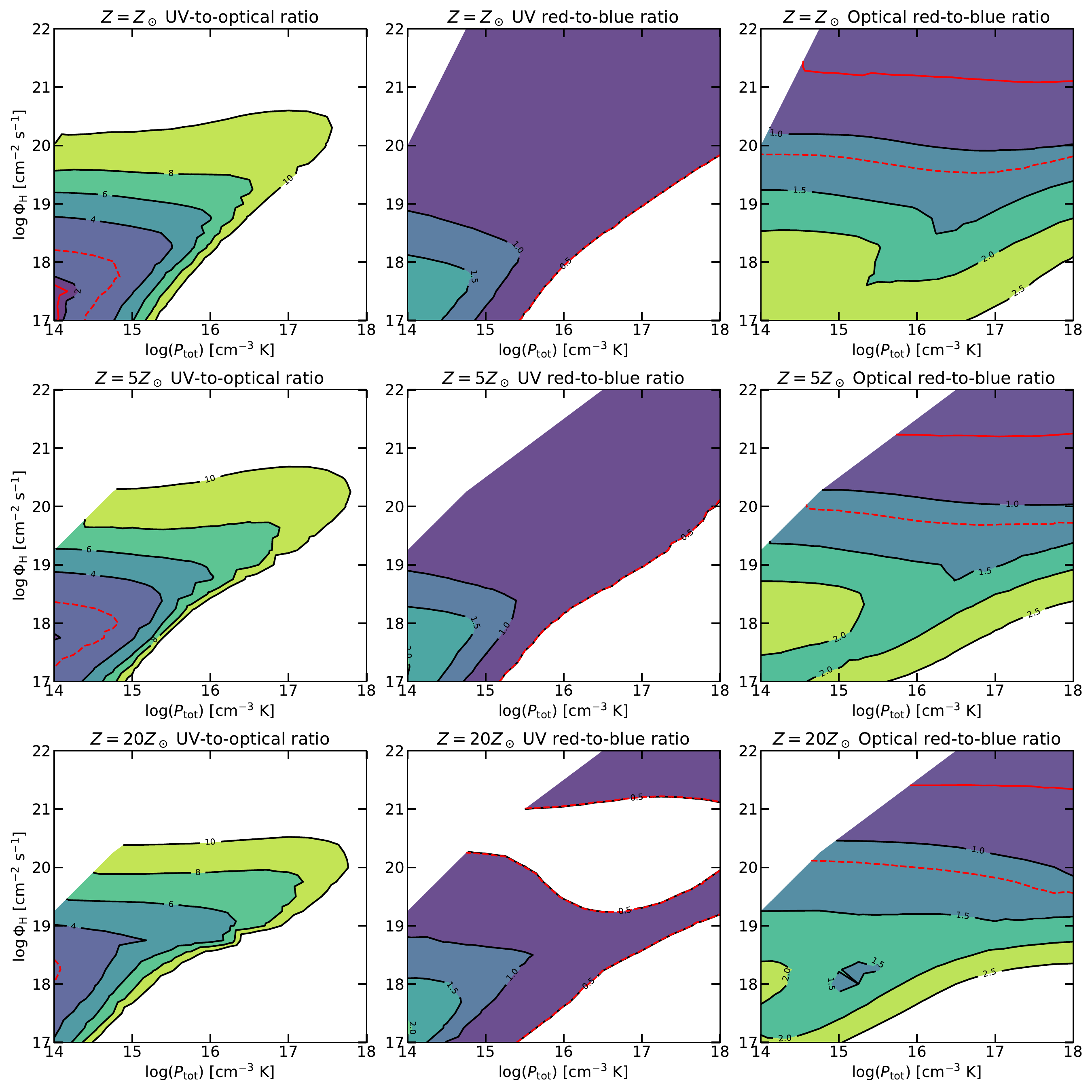}
    \caption{Contour plots for the canonical ratios described in Section \ref{sect:fitting} with constant $V_{\rm turb}= 0$ km~s$^{-1}$ and assuming equal contribution from bright and dark sides of clouds. {\it Top row}: Contour plots for the described ratios for $Z=Z_\odot$. {\it Middle row}: Contour plots for the described ratios for $Z=5Z_\odot$. {\it Bottom row}: Contour plots for the described ratios for $Z=20Z_\odot$. For each case, the position occupied in the parameter space by the measured corresponding ratio is displayed by a red line, while contours associated with the 1-$\sigma$ confidence range of the ratio are characterized by a dashed red line.}
    \label{fig:threeratio_trends0}
\end{figure*}

\begin{figure*}[h]
    \centering
    \includegraphics[width=18.5cm]{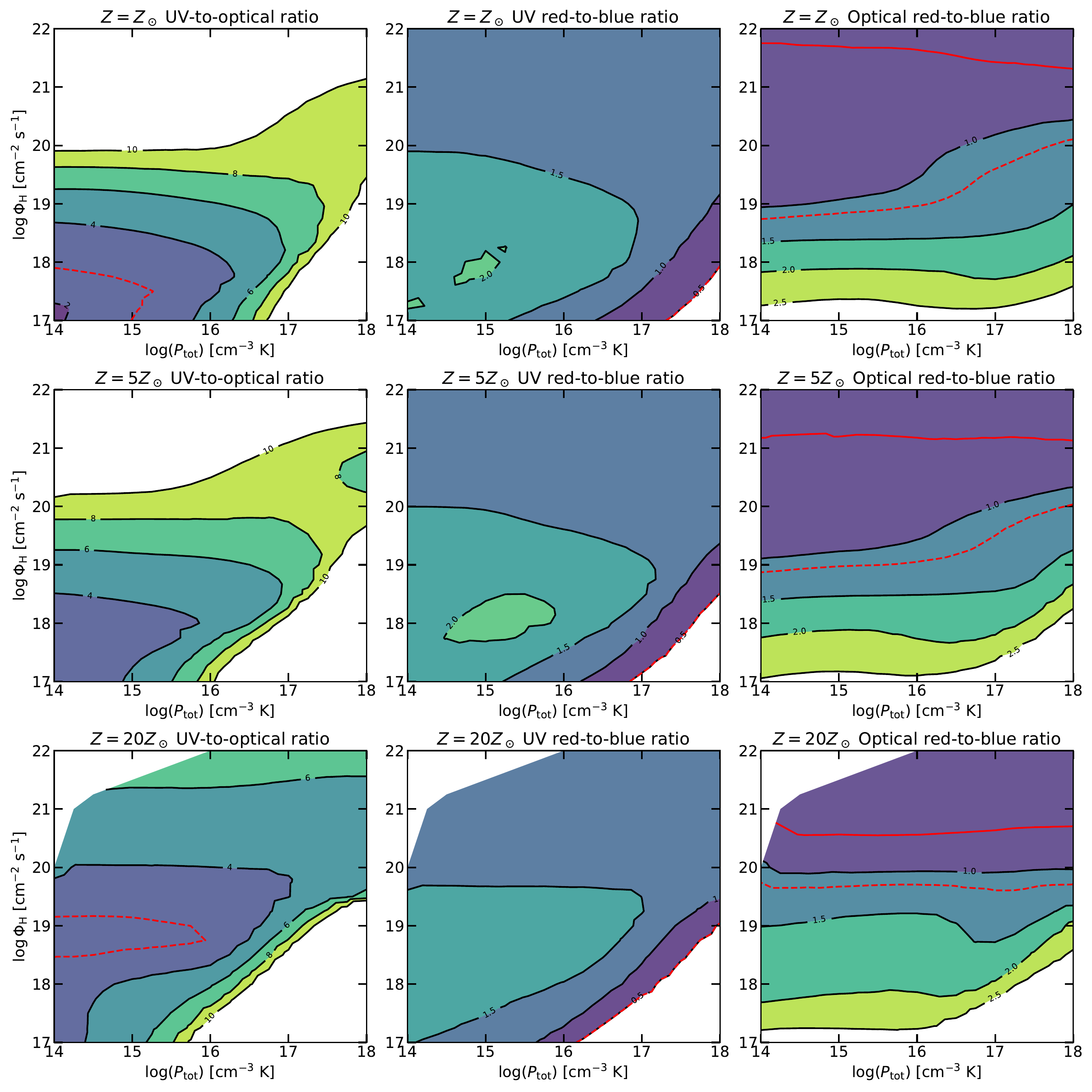}
    \caption{Same as in Fig. \ref{fig:threeratio_trends0}, but for $V_{\rm turb}= 40$ km~s$^{-1}$.}
    \label{fig:threeratio_trends40}
\end{figure*}

\begin{figure*}[h]
    \centering
    \includegraphics[width=18.5cm]{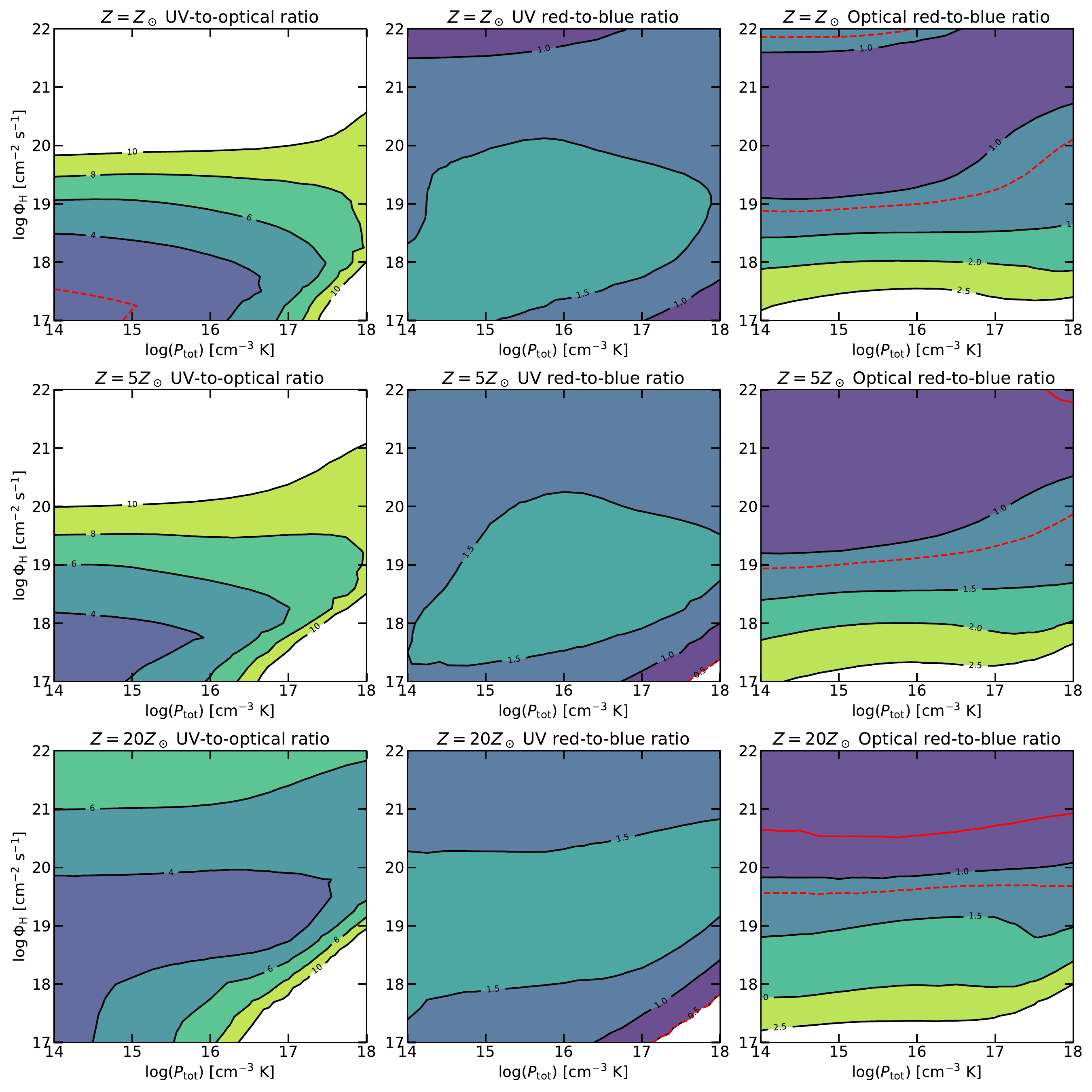}
    \caption{Same as in Fig. \ref{fig:threeratio_trends0}, but for $V_{\rm turb}= 100$ km~s$^{-1}$.}
    \label{fig:threeratio_trends100}
\end{figure*}

\begin{figure*}[h]
    \centering
    \includegraphics[width=18.5cm]{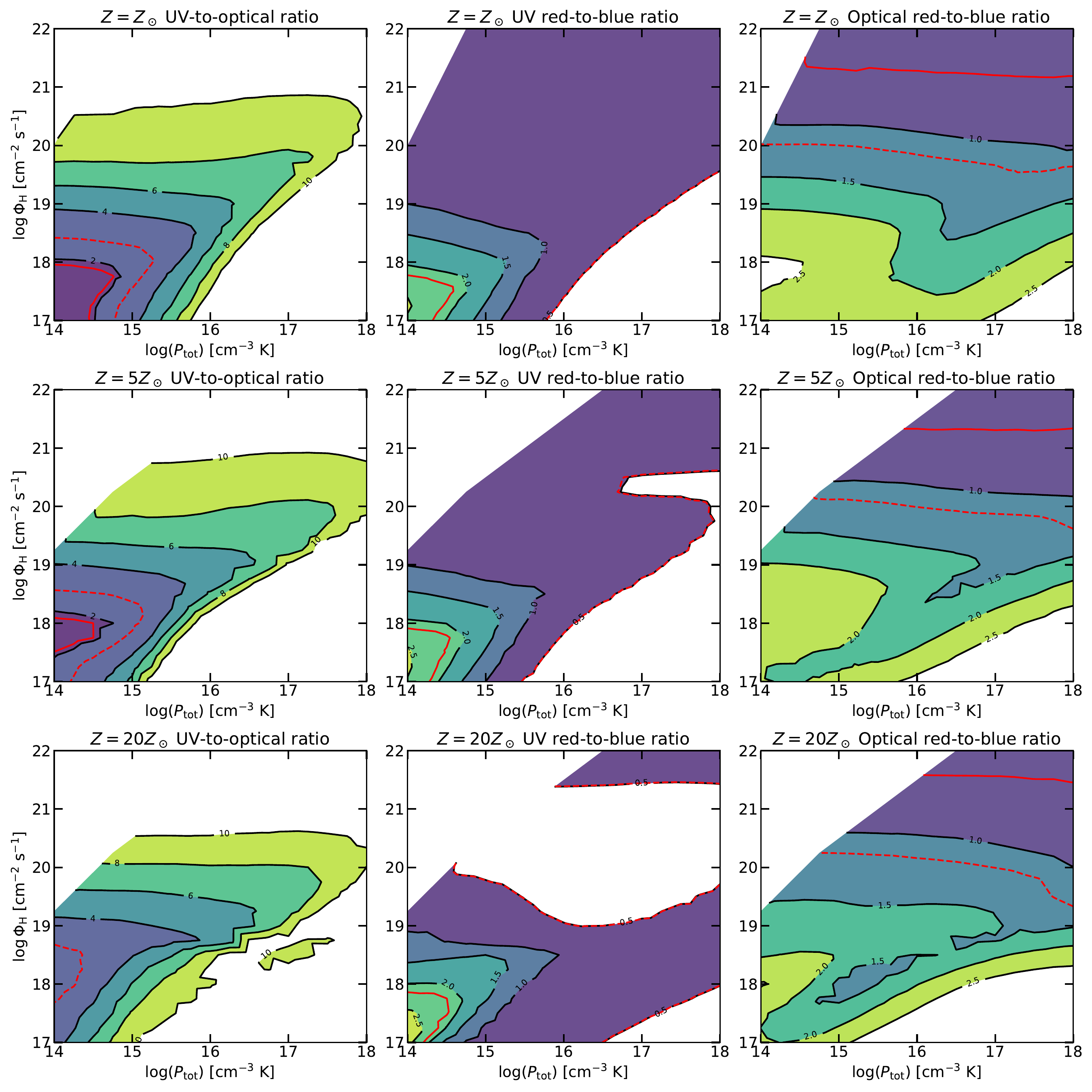}
    \caption{Same as in Fig. \ref{fig:threeratio_trends0}, but assuming 20\% contribution from bright sides and 80\% contribution from dark sides of clouds.}
    \label{fig:threeratio_trends0_2080}
\end{figure*}

\begin{figure*}[h]
    \centering
    \includegraphics[width=18.5cm]{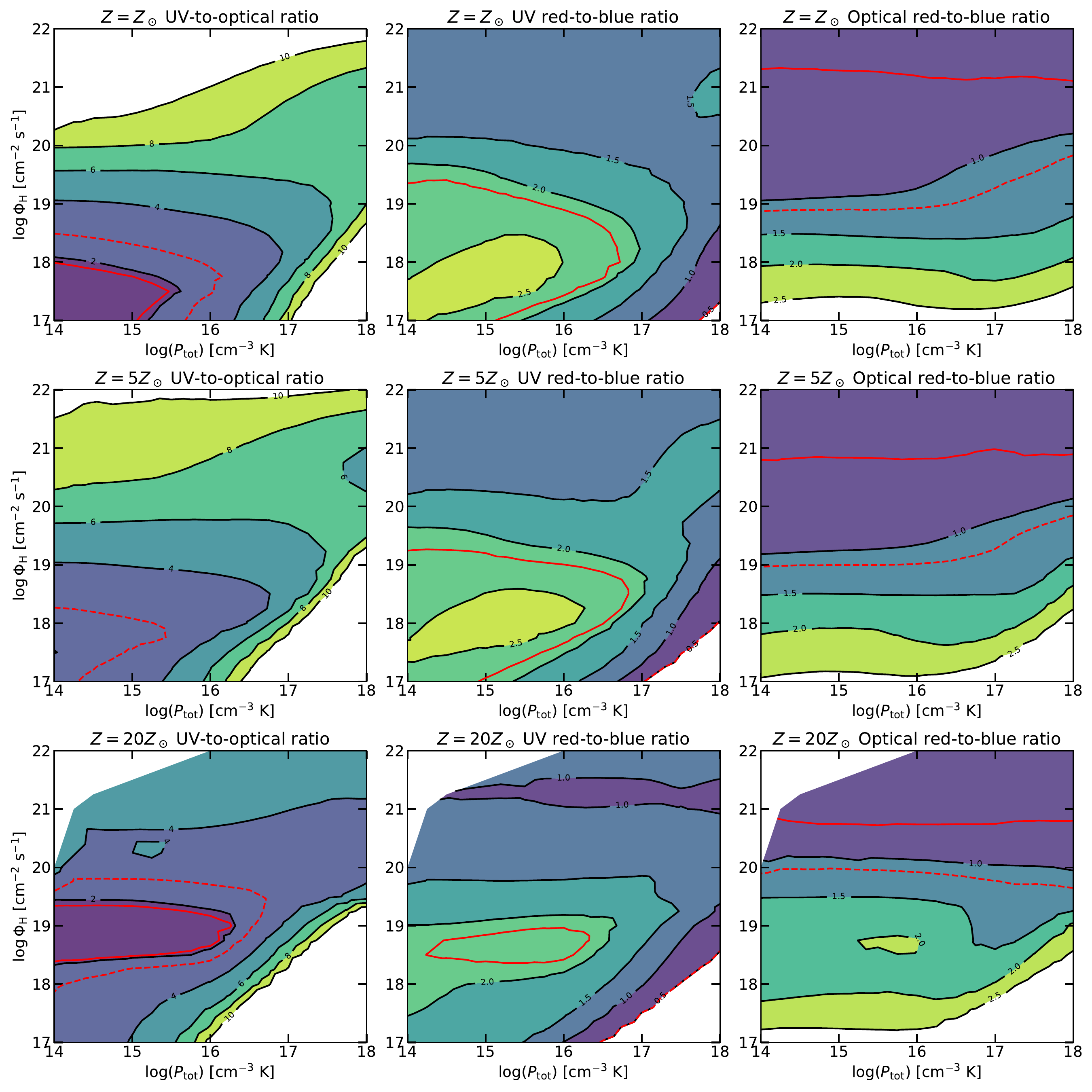}
    \caption{Same as in Fig. \ref{fig:threeratio_trends0}, but for $V_{\rm turb}= 40$ km~s$^{-1}$ and assuming 20\% contribution from bright sides and 80\% contribution from dark sides of clouds.}
    \label{fig:threeratio_trends40_2080}
\end{figure*}

\begin{figure*}[h]
    \centering
    \includegraphics[width=18.5cm]{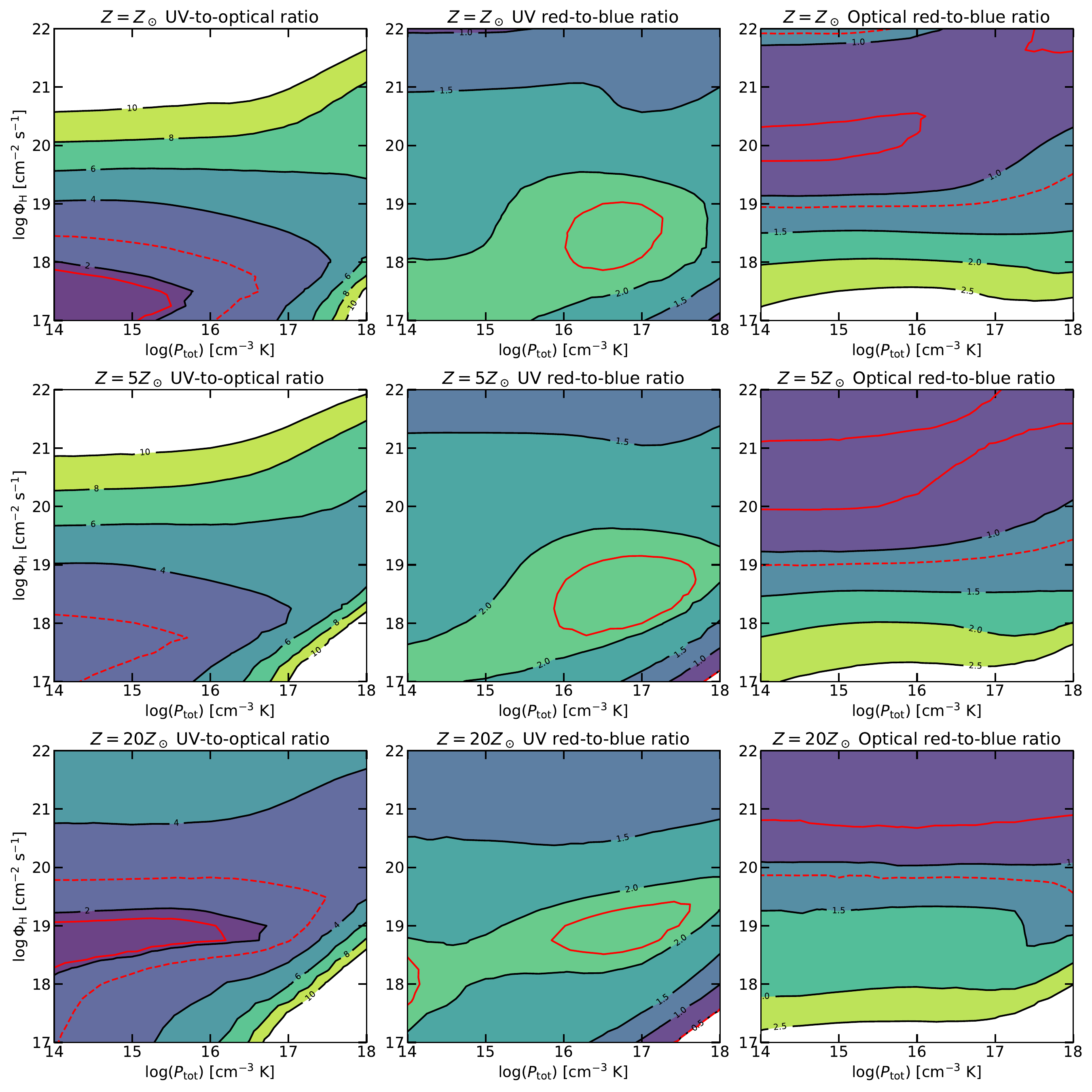}
    \caption{Same as in Fig. \ref{fig:threeratio_trends0}, but for $V_{\rm turb}= 100$ km~s$^{-1}$ and assuming 20\% contribution from bright sides and 80\% contribution from dark sides of clouds.}
    \label{fig:threeratio_trends100_2080}
\end{figure*}

From Figures \ref{fig:threeratio_trends0} to \ref{fig:threeratio_trends100}, we notice that, assuming equal contribution from illuminated and dark sides, the observed UV-to-optical ratio for RM 102 is rarely reproduced, except in simulations with very low $n_{\rm H}$ and low $\Phi$(H). The contours of the ratio mostly retain the shape with changing microturbulence parameter and metallicity. When considering a 20\% contribution from the illuminated sides and 80\% from the dark sides, the observed UV-to-optical ratio is more frequently produced at moderate values of $n_{\rm H}$ and $\Phi$(H), especially with increasing $Z$ and $V_{\rm turb}$. Values within the error of the ratio are consistently reproduced in much of the parameter space when $V_{\rm turb} = 100$ km~s$^{-1}$ and $Z=20Z_\odot$, strongly suggesting this configuration as the most likely for the physical conditions in the BLR of RM 102.


\subsubsection{UV red-to-blue ratio}

Similarly to the UV-to-optical ratio, assuming equal contributions from illuminated and dark sides, the observed value of the UV red-to-blue ratio is never produced, with the ratio generally being much lower across the studied parameter space. This ratio, which is influenced by numerous UV {\rm Fe{\sc ii}} transitions around {\rm Mg{\sc ii}}$\lambda2800$, follows a complicated trend and is underproduced in simulations with high $n_{\rm H}$ and $\Phi$(H). However, when considering a 20\% contribution from illuminated sides and 80\% from dark sides, the ratio is more consistently produced across all values of $V_{\rm turb}$ and $Z$. In particular, for $V_{\rm turb} > 10$ km~s$^{-1}$, the observed ratio is produced in simulations for combinations of moderate $P_{\rm tot}$ and $\Phi$(H). The ratio shows little dependence on metallicity, and contour shapes remain stable with increasing microturbulence for $V_{\rm turb} > 10$ km~s$^{-1}$. 

\subsubsection{Optical red-to-blue ratio}

The optical red-to-blue ratio behaves differently from the other two ratios. The observed ratio for RM 102 is consistently produced across all cases studied, regardless of whether there is a 50-50 or 20-80 contribution from bright and dark sides. The ratio mainly depends on the value of $\Phi$(H) and is generally independent of other parameters, with a few exceptions when $V_{\rm turb} = 100$ km~s$^{-1}$ and when there is a 20\% contribution from bright sides and 80\% from dark sides. In these cases, the ratio follows a more complex trend, appearing at $\Phi$(H) values compatible with the measured ionizing flux of the source for $Z$=1-5$Z_\odot$. However, in most of the simulations, the predicted position occupied by the ratio in the parameter space is inconsistent with the measured $\Phi$(H) in the BLR of the object, indicating a potential unresolved inconsistency.


\subsection{Trends of additional ratios}
\label{app:other_ratios}

\begin{figure*}[h]
    \centering
    \includegraphics[width=18.5cm]{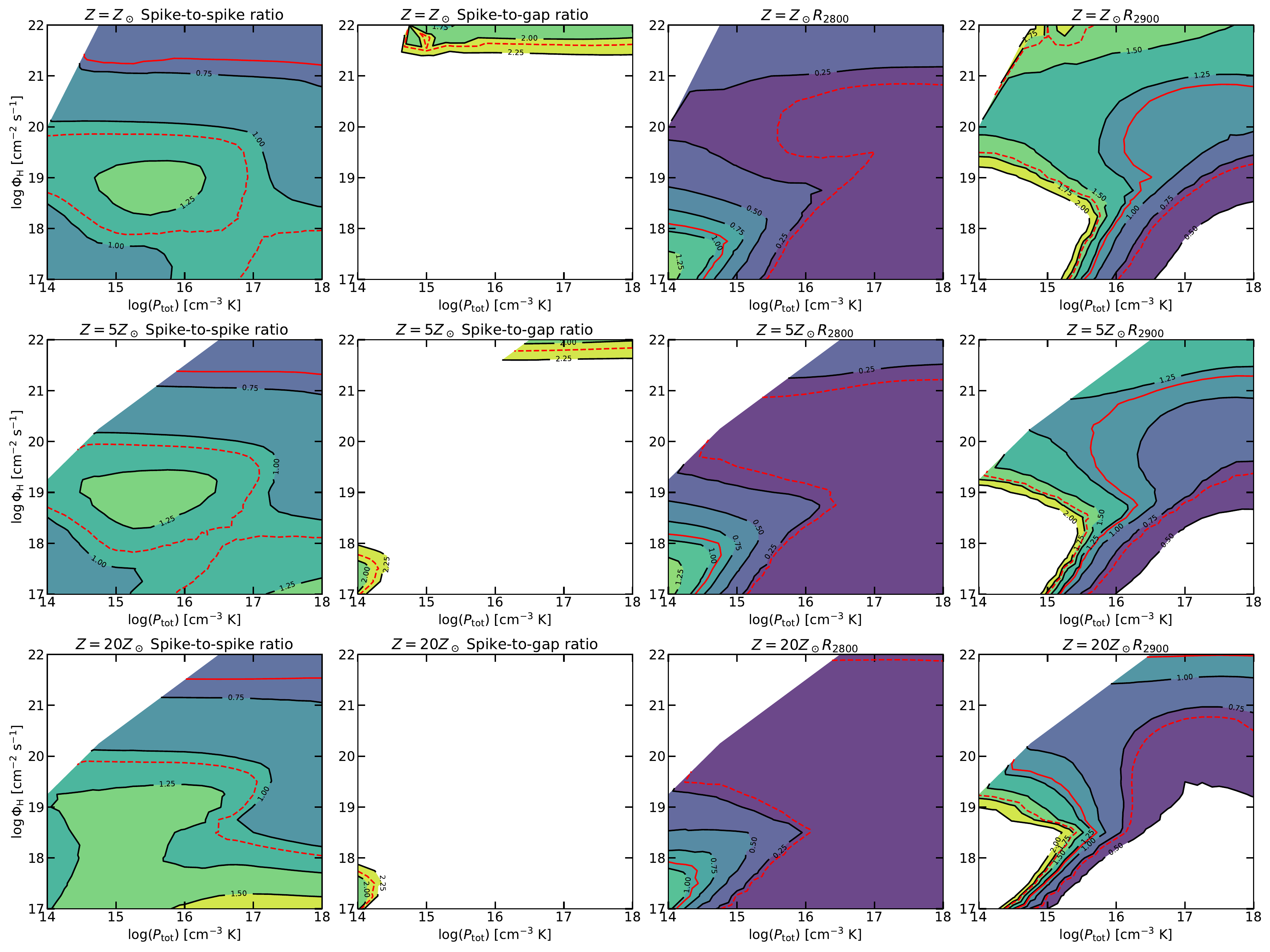}
    \caption{Contour plots for the ratios discussed in Section \ref{transitions} with constant $V_{\rm turb}= 0$ km~s$^{-1}$ and assuming equal contribution from bright and dark sides of clouds. {\it Top row}: Contour plots for the described ratios for $Z=Z_\odot$. {\it Middle row}: Contour plots for the described ratios for $Z=5Z_\odot$. {\it Bottom row}: Contour plots for the described ratios for $Z=20Z_\odot$. For each case, the position occupied in the parameter space by the measured corresponding ratio is displayed by a red line.}
    \label{fig:Ratio_trends0}
\end{figure*}

\begin{figure*}[h]
    \centering
    \includegraphics[width=18.5cm]{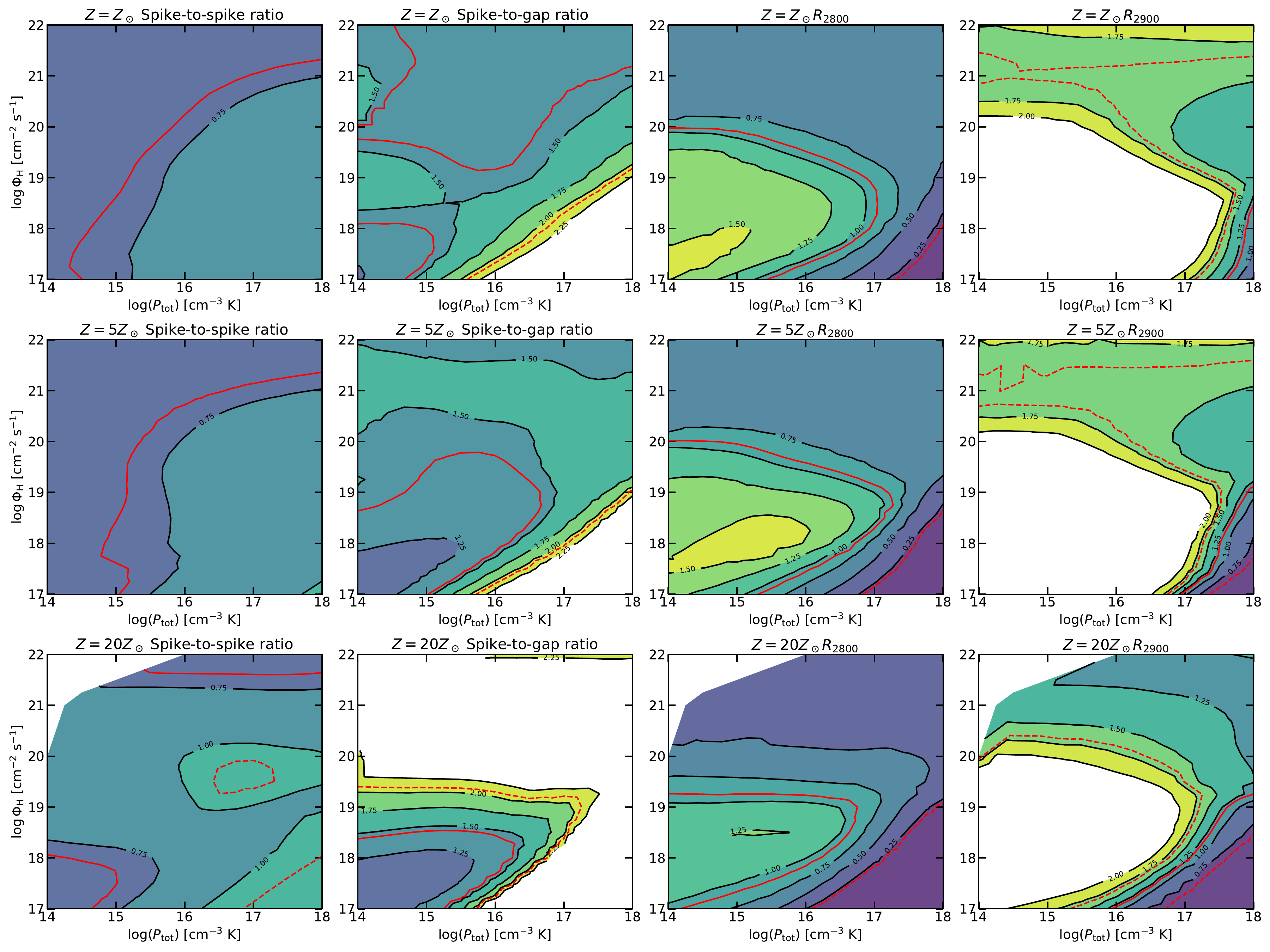}
    \caption{Same as in Fig. \ref{fig:Ratio_trends0}, but for $V_{\rm turb}= 40$ km~s$^{-1}$.}
    \label{fig:Ratio_trends40}
\end{figure*}

\begin{figure*}[h]
    \centering
    \includegraphics[width=18.5cm]{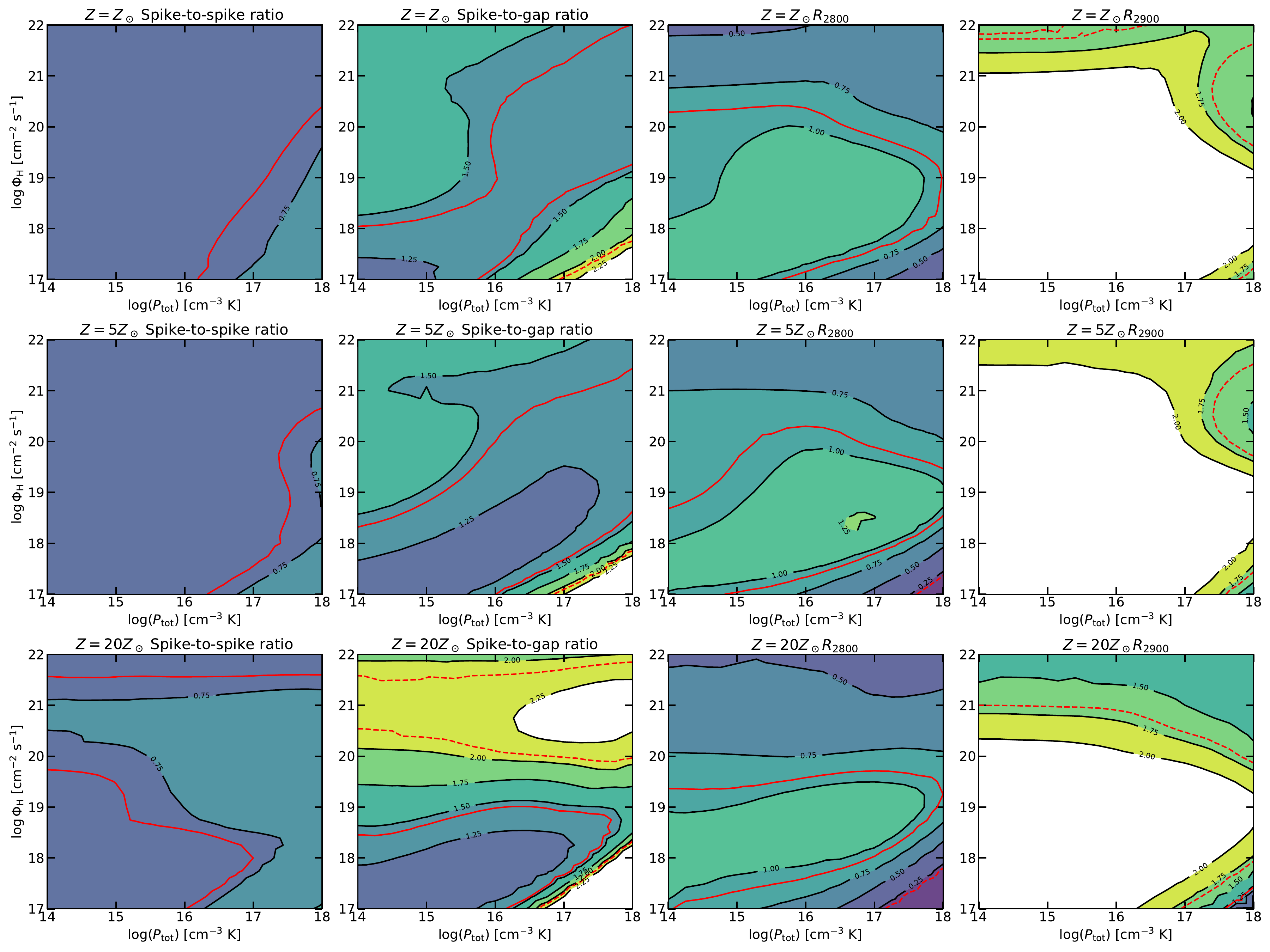}
    \caption{Same as in Fig. \ref{fig:Ratio_trends0}, but for $V_{\rm turb}= 100$ km~s$^{-1}$.}
    \label{fig:Ratio_trends100}
\end{figure*}

\begin{figure*}[h]
    \centering
    \includegraphics[width=18.5cm]{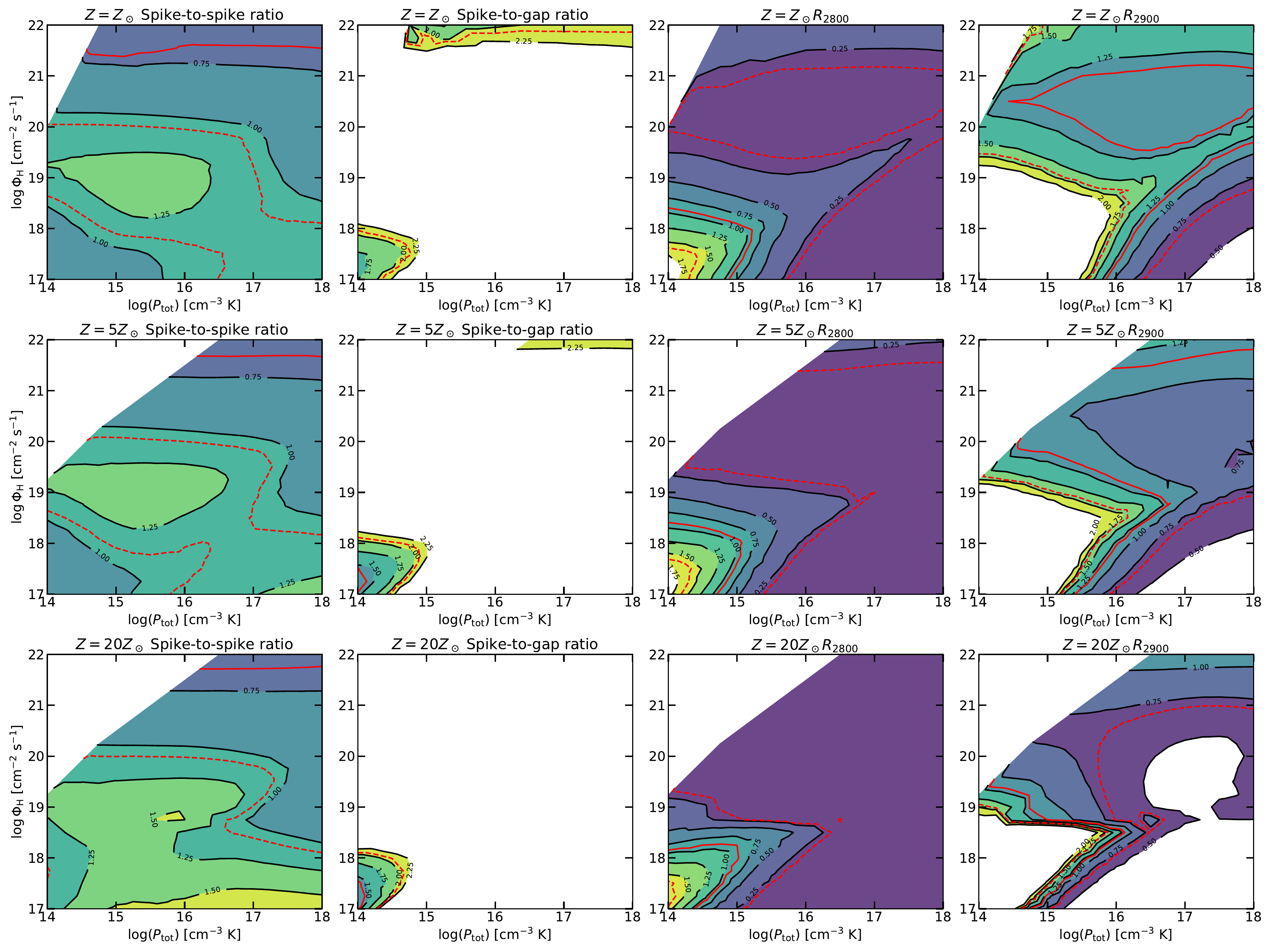}
    \caption{Same as in Fig. \ref{fig:Ratio_trends0}, but assuming 20\% contribution from bright sides and 80\% contribution from dark sides of clouds.}
    \label{fig:Ratio_trends0_2080}
\end{figure*}

\begin{figure*}[h]
    \centering
    \includegraphics[width=18.5cm]{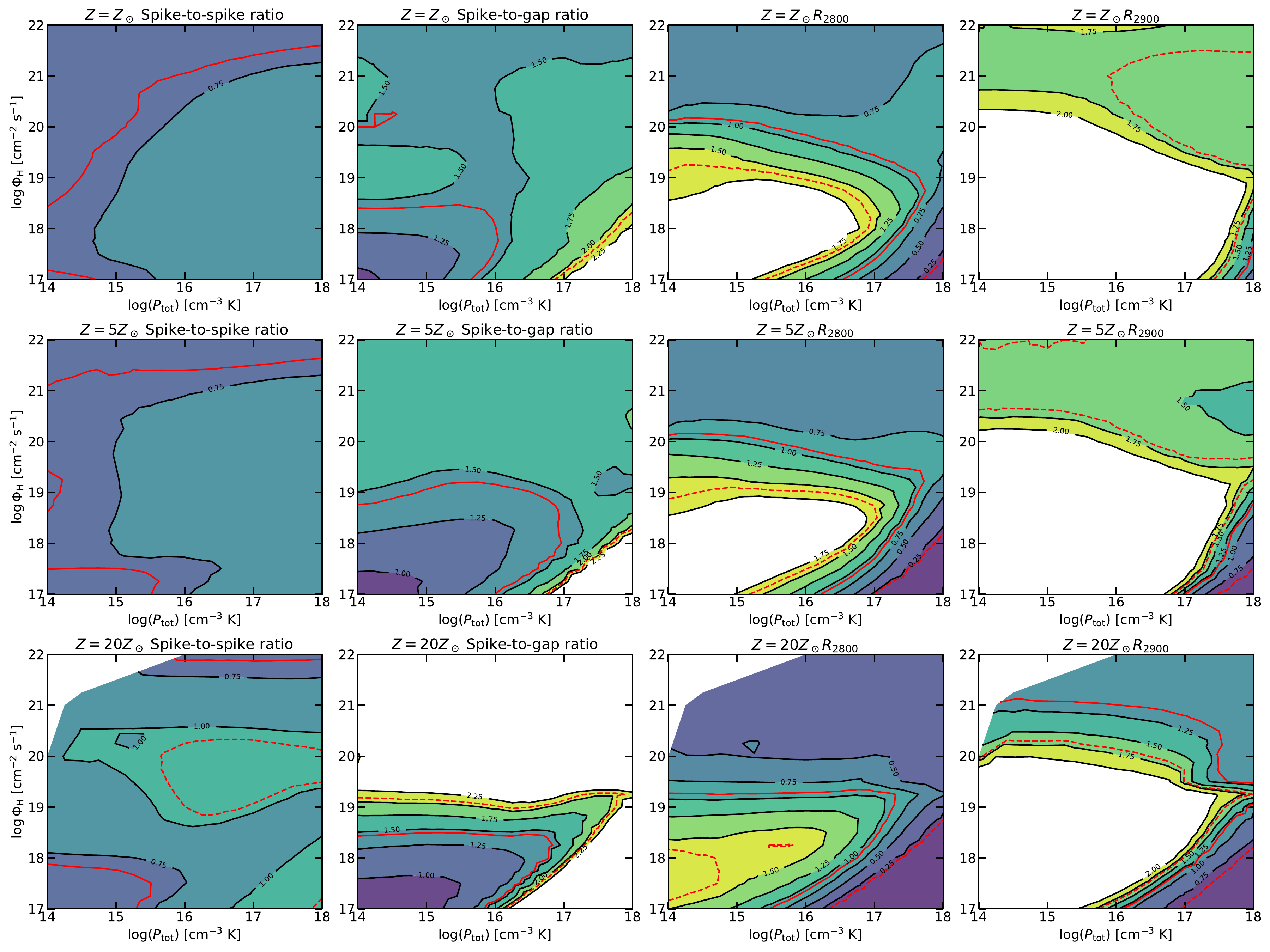}
    \caption{Same as in Fig. \ref{fig:Ratio_trends0}, but for $V_{\rm turb}= 40$ km~s$^{-1}$ and assuming 20\% contribution from bright sides and 80\% contribution from dark sides of clouds.}
    \label{fig:Ratio_trends40_2080}
\end{figure*}

\begin{figure*}[h]
    \centering
    \includegraphics[width=18.5cm]{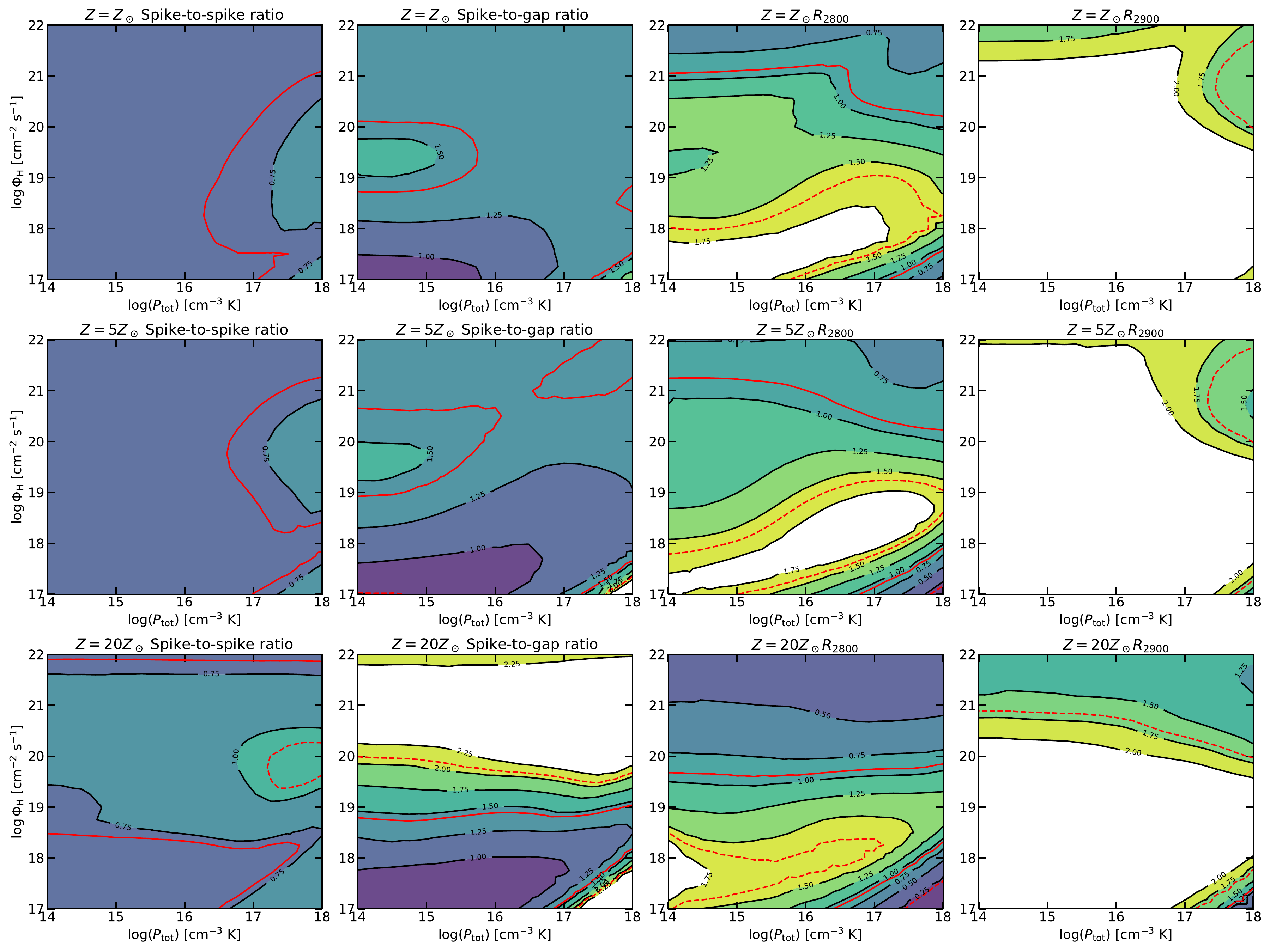}
    \caption{Same as in Fig. \ref{fig:Ratio_trends0}, but for $V_{\rm turb}= 100$ km~s$^{-1}$ and assuming 20\% contribution from bright sides and 80\% contribution from dark sides of clouds.}
    \label{fig:Ratio_trends100_2080}
\end{figure*}

Figures \ref{fig:Ratio_trends0}-\ref{fig:Ratio_trends100_2080} display the trends associated with the additional ratios defined in Section \ref{transitions} within the selected parameter space of $\log\Phi$(H), $\log(P_{\rm tot})$, $V_{\rm turb}$ and $Z$. 

\subsubsection{Spike-to-spike ratio}

The spike-to-spike ratio observed in RM 102 serves as a key diagnostic for high microturbulence. In the absence of microturbulence, the observed ratio is reproduced only at extreme values of $\Phi$(H) which are incompatible with the observations. Conversely, for $V_{\rm turb}$ = 40-100 km~s$^{-1}$, the ratio is consistently produced for $Z=$ 1-5 $Z_\odot$ under typical BLR conditions compatible with the observed $\Phi$(H). At low microturbulence, the ratio primarily depends on $\Phi$(H) at high $\Phi$(H), while following a more complex pattern at low ionizing flux and high density. The spike-to-spike ratio is largely independent of whether equal or unequal contributions from dark and illuminated sides are considered.

\subsubsection{Spike-to-gap ratio}

The spike-to-gap ratio, as also outlined in \cite{sarkar2021}, mainly depends on the microturbulence parameter. The typical value of $\sim 1.4$ is almost never reached in simulations in absence of microturbulence. For simulations with $V_{\rm turb}= 0$ km~s$^{-1}$, the minimum predicted value is around 2, appearing only for extreme values of $\Phi$(H) and $n_{\rm H}$. For $V_{\rm turb}>$ 20 km~s$^{-1}$, the ratio is consistently reproduced at moderate values of $\Phi$(H) and $P_{\rm tot}$. In this case, the best-fitting solution appears to correspond to $Z\sim5Z_\odot$. A strong dependence on metallicity is noted, with the spikes being most efficiently produced for increasing values of $Z$ for typical values of $\Phi$(H) and $n_{\rm H}$. The ratio remains consistent regardless of whether equal or unequal contributions from bright and dark sides are considered.

\subsubsection{$R_{2800}$ ratio}

The $R_{2800}$ ratio follows an opposite trend compared to the spike-to-gap ratio, typically decreasing with higher values of $\Phi$(H) and $n_{\rm H}$. In the absence of microturbulence, the parameter space is mainly occupied by ratio values smaller than 0.25, while the observed $R_{2800}$ ratio for RM 102 is 0.909. With increasing $V_{\rm turb}$, the ratio is more consistently reproduced under conditions compatible with the observations.
The ratio does not strongly depend on $Z$, as contours retain their shape across different $Z$ values. With increasing microturbulence, the observed ratio is predicted for more moderate values of $\Phi$(H) and $n_{\rm H}$.
For $V_{\rm turb}=100$ km~s$^{-1}$, the ratio becomes mainly independent of $P_{\rm tot}$. The regions of the parameter space compatible with the observed ratio remain largely unchanged whether considering equal or unequal contributions from bright and dark sides.

\subsubsection{$R_{2900}$ ratio}

The $R_{2900}$ ratio is highly sensitive to variations in metallicity, especially at high $V_{\rm turb}$.
There is a progressive clustering of ratio values at low $\Phi$(H) and $n_{\rm H}$, with the ratio rapidly diverging toward extreme values. With increasing microturbulence, the observed ratio is predicted only under high-density conditions, although the interpretation becomes more complex at $Z=20Z_\odot$. When considering unequal contributions from bright and dark sides, the clustering becomes even more pronounced.

\section{Equivalent width trends}
\label{ew}

\begin{figure*}[h]
    \centering
    \includegraphics[width=18.5cm]{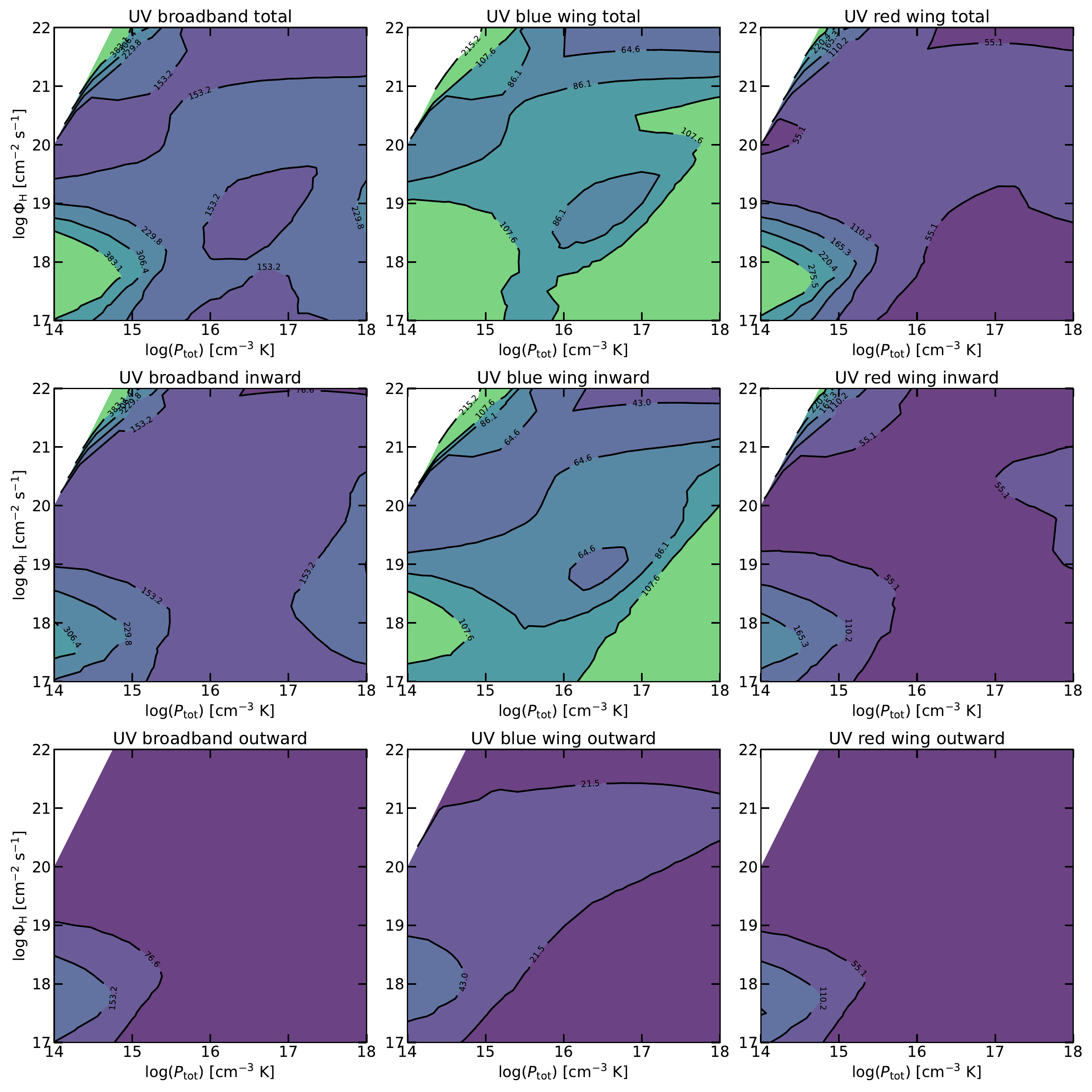}
    \caption{Contour plots for the predicted EW from {\tt CLOUDY} simulations for the case of solar metallicity and no turbulent velocity. The plots show contours for EWs corresponding to covering factors of $f=1$, $f=0.5$, $f=0.33$, $f=0.25$, $f=0.2$ and $f=0.1$.}
    \label{fig:contourewz1t0}
\end{figure*}

\begin{figure*}[h]
    \centering
    \includegraphics[width=18.5cm]{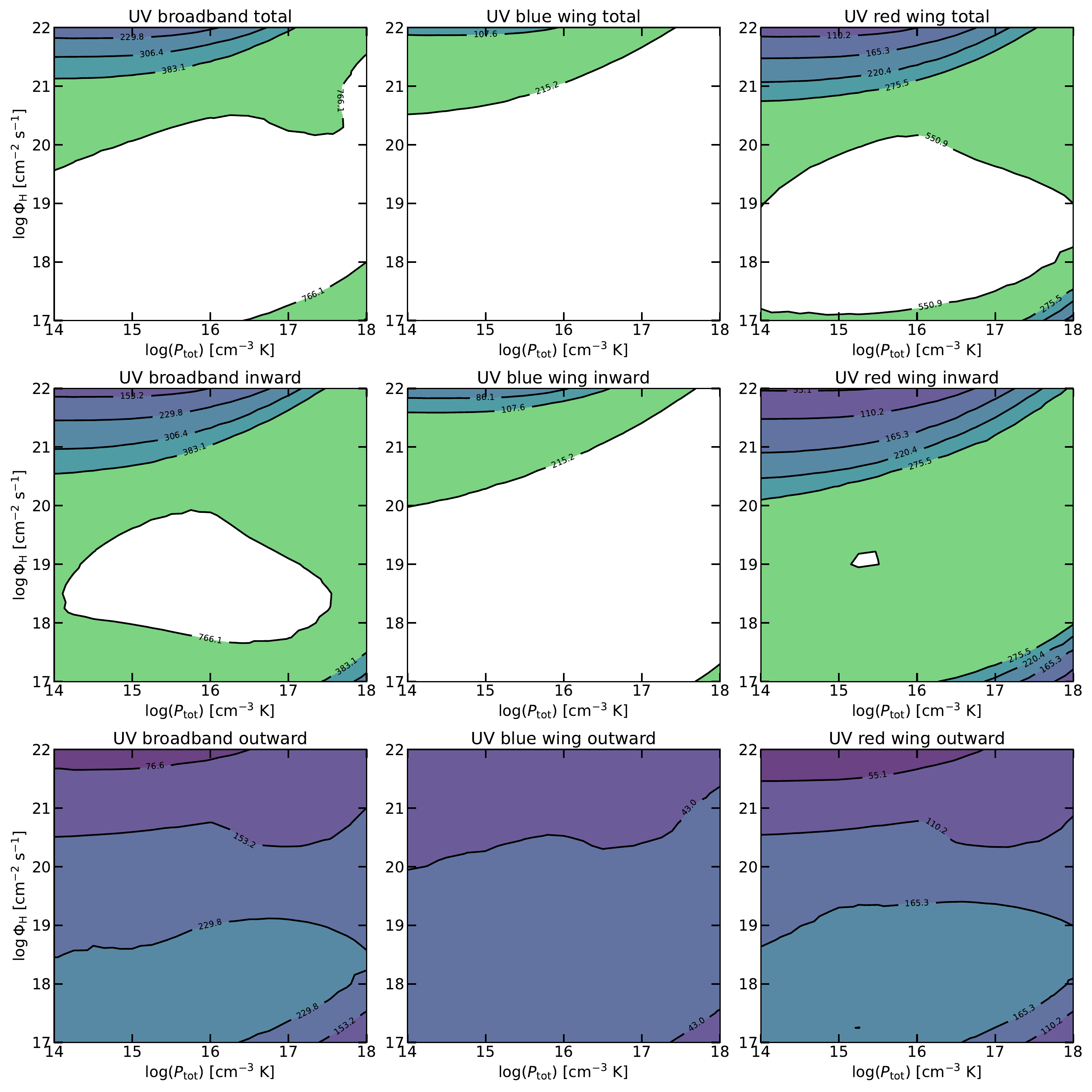}
    \caption{Same as in Fig. \ref{fig:contourewz1t0}, but for $Z=Z_\odot$ and $V_{\rm turb}=100$ km s$^{-1}$.}
    \label{fig:contourewz1t100}
\end{figure*}

\begin{figure*}[h]
    \centering
    \includegraphics[width=18.5cm]{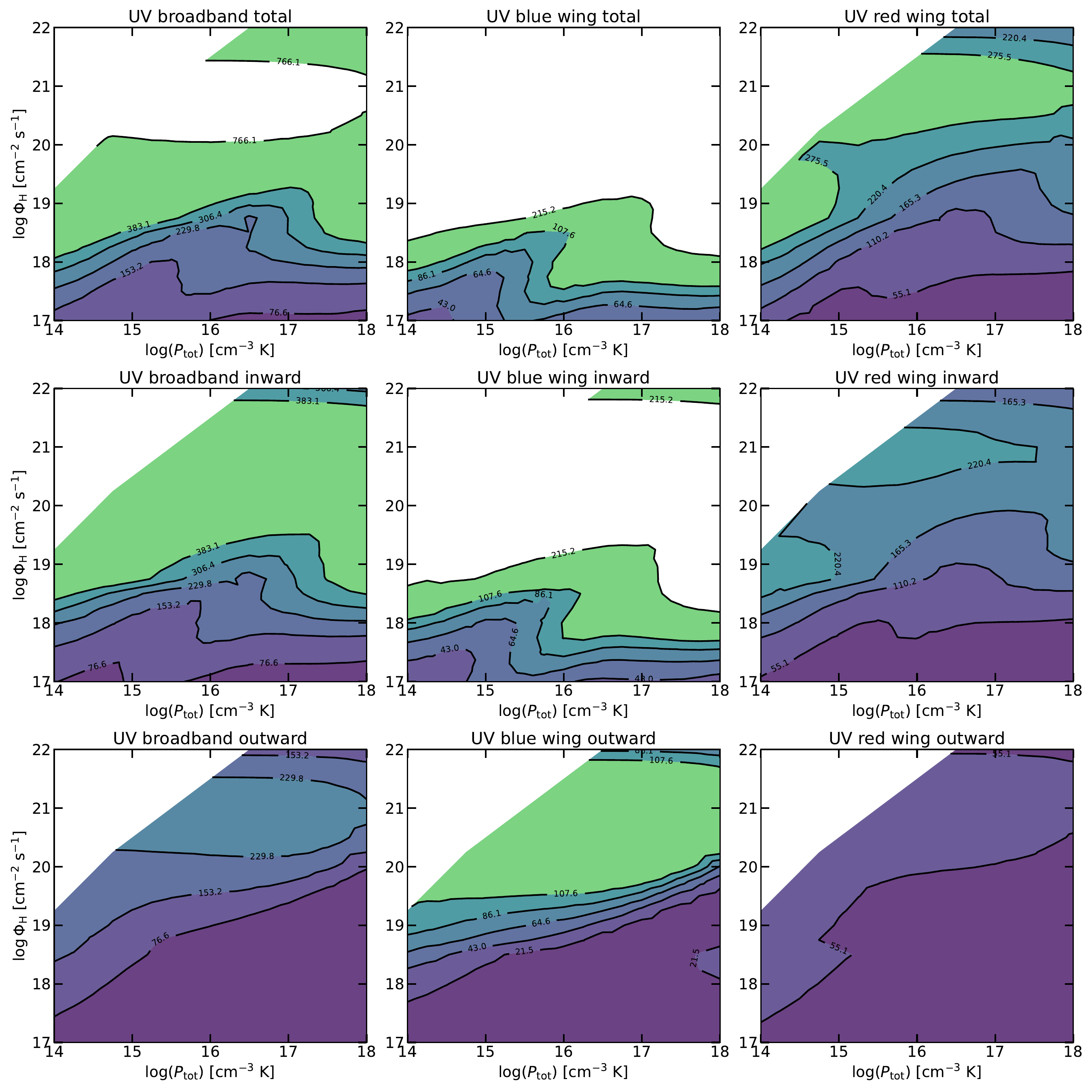}
    \caption{Same as in Fig. \ref{fig:contourewz1t0}, but for $Z=20Z_\odot$ and $V_{\rm turb}=0$ km s$^{-1}$.}
    \label{fig:contourewz20t0}
\end{figure*}

\begin{figure*}[h]
    \centering
    \includegraphics[width=18.5cm]{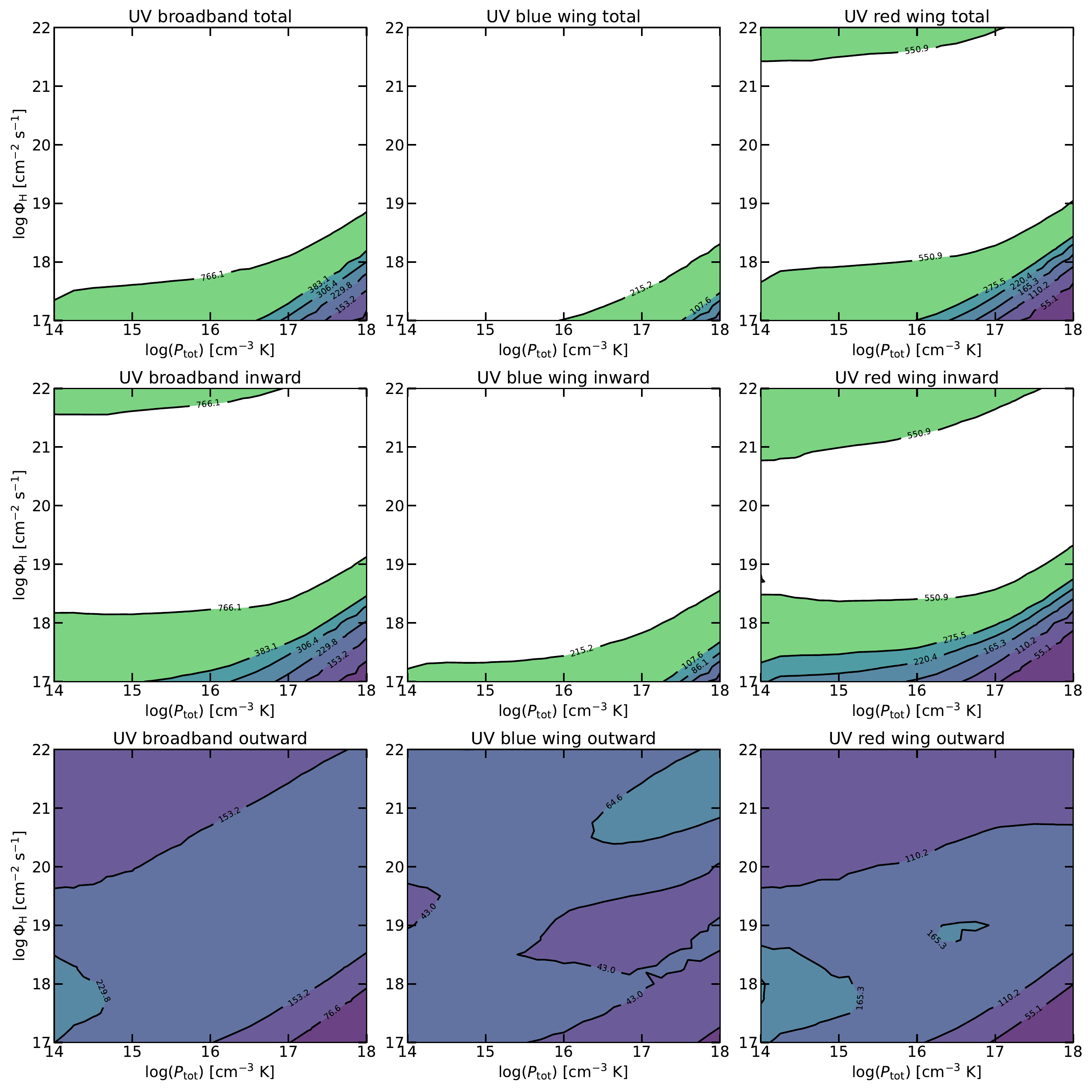}
    \caption{Same as in Fig. \ref{fig:contourewz1t0}, but for $Z=20Z_\odot$ and $V_{\rm turb}=100$ km s$^{-1}$.}
    \label{fig:contourewz20t100}
\end{figure*}

Figures \ref{fig:contourewz1t0}-\ref{fig:contourewz20t100} display the trends associated with the {\rm Fe{\sc ii}} EW production from the simulations considered in this work. Each plot illustrates the EW produced in both the blue and red UV wings, along with the broadband emission, and the inward and outward radiation emitted from the illuminated and dark sides of the clouds, respectively. These exemplary cases are shown for for different combinations of $V_{\rm turb}$ and $Z$. Different contours correspond to various values of the covering factor adopted to match the {\rm Fe{\sc ii}} emission of RM 102.

\end{appendix}

%
%

\end{document}